\documentclass[fleqn,usenatbib]{mnras}

\usepackage[switch]{lineno}   
\usepackage{newtxtext,newtxmath}
\usepackage{graphicx}
\usepackage{subcaption}
\usepackage{lipsum}

\usepackage[T1]{fontenc}

\DeclareRobustCommand{\VAN}[3]{#2}
\let\VANthebibliography\thebibliography
\def\thebibliography{\DeclareRobustCommand{\VAN}[3]{##3}\VANthebibliography}

\usepackage{graphicx}	
\usepackage{amsmath}	

\title[Is accretion flow magnetically arrested in M87?]{Is the accretion flow magnetically arrested in M87?}

\author[Ma et al.]{
Yuao Ma$^{1}$,
Xinwu Cao$^{1,2}$\thanks{E-mail: xwcao@zju.edu.cn (XC)},
Jia-Wen Li$^{3}$\thanks{E-mail: jwliynu@ynu.edu.cn (JL)}, 
Jianchao Feng$^{4}$, and  
Qingwen Wu$^5$
\\
$^{1}$Institute for Astronomy, School of Physics, Zhejiang University, 866 Yuhangtang Road, Hangzhou 310058, China\\
$^{2}$Center for Cosmology and Computational Astrophysics, School of Physics, Zhejiang University, 866 Yuhangtang Road, Hangzhou 310058, China\\
$^{3}$Department of Astronomy, School of Physics and Astronomy, Key Laboratory of Astroparticle Physics of Yunnan Province, Yunnan University, \\Kunming 650091, People's Republic of China\\
$^{4}$School of Physics and Electronic Science, Guizhou Normal University, Guiyang 550001, China\\
$^{5}$Department of Astronomy, School of Physics, Huazhong University of Science and Technology, Luoyu Road 1037, Wuhan, People's Republic of China
}

\date{Accepted 2026 August 28. Received 2026 August 06; in original form 2026 April 22}

\pubyear{2015}

\begin{document}
\label{firstpage}
\pagerange{\pageref{firstpage}--\pageref{lastpage}}
\maketitle

\begin{abstract}

It is still debating on whether the disc is magnetically arrested in M87.  We assume that a weak external magnetic field is dragged inwards by the accretion disc, which is substantially enhanced to drive strong jets near the black hole horizon via Blandford-Znajek mechanism. The jet power of M87 has been well constrained with the observational data, while the accretion rate in the inner region of the accretion flow in M87 is estimated by fitting the multi-waveband continuum spectrum and the data of the Faraday rotation measurement, with which the surface density of the disc is derived.  Our calculations show that, in order to produce the observed jet power $P_{\rm jet}=3.2\times 10^{43}~\rm erg~s^{-1}$, an external field with several hundred $\mu \rm G$ at the outer edge of the disc is required to be amplified in the disc to hundreds G at the BH horizon,  and the accretion flow in M87 must be magnetically arrested.  A standard and normal evolution (SANE) disc is allowed in M87, only if the jet power is significantly lower than $\sim 1.75 \times 10^{43} \rm erg~ s^{-1}$,

\end{abstract}




\section{Introduction}\label{sec_intro}

M87 (Messier 87) is an elliptical galaxy located near the center of the Virgo Cluster, approximately $16.8 \pm 0.8 \rm Mpc$ away from Earth \citep{2009ApJ...694..556B}. Its most notable feature is the presence of a supermassive black hole (BH) at its center, with a mass approximately $(6.5\pm0.7)\times10^9 M_{\odot}$ \citep{2019ApJ...875L...6E}, derived from its central stellar dynamics. This BH is not only the subject of the first-ever image of a BH captured by the Event Horizon Telescope (EHT), but also drives a relativistic jet of extremely high energy that extends over $60 \rm kpc$ \citep{2021ApJ...911L..11E}. 

Similar to the stellar-mass BHs in X-ray binaries, there are different accretion modes for supermassive BHs in active galactic nuclei (AGNs). It is believed that geometrically  thin cold accretion discs are surrounding BHs in luminous AGNs, while the central BHs in low-luminosity AGNs or normal galaxies are surrounded by geometrically thick hot accretion flows. Indeed, such hot accretion flows in M87 and the centre of our own Galaxy have been imaged with EHT.  The properties of the accretion flow in M87 can be derived with the advection dominated accretion flow (ADAF) model \citep{1995ApJ...452..710N,2016ApJ...830....6F,2017MNRAS.470..612F,2024ApJ...976..214N}. \cite{2016ApJ...830....6F} found that the millimeter bump observed with the Atacama Large Millimeter/submillimeter Array (ALMA) can be well-modeled by synchrotron emission from thermal electrons within an ADAF,   
which is also consistent with the Very Long Baseline Interferometry observations conducted at $230 \rm GHz$. Their estimate of the accretion rate close to the BH horizon is $\dot{M}\sim (0.2-1)\times 10^{-3}M_\odot~ \rm yr^{-1}$ with the SED modeling and the Faraday rotation measurements.

M87 is a Fanaroff-Riley (FR) I radio galaxy with kiloparsec-scale jets, which exhibit significant emission in the radio, optical, near-ultraviolet (NUV), and X-ray bands \citep{2018ApJ...855..128W}. 
The power of the jets in M87 galaxy is a classic case for studying the feedback effect of active galactic nuclei. The jet power plays an important role in AGN feedback, however, it is still a difficult mission to measure it accurately.  There are a pair of lobes/cavities observed in this source, and one can estimate the energy stored in the lobes/cavities through multi-band (rado or/and X-ray) observations. The expanding timescale to form such cavities can be estimated with somewhat model dependent method, and the jet power averaged over the expanding timescale is derived \citep[e.g.,][and the references therein]{2006MNRAS.372...21A} .  The estimates of the jet power of M87 disperse in a large range of   $10^{42} - 10^{44} \rm erg~s^{-1}$ \citep{2002ApJ...579..560Y,2006MNRAS.372...21A,2012A&A...547A..56D}. 
As the jets emit predominantly in radio wave-bands, the core radio emission reflects the jet power more directly, which is the instant jet power. \citet{1998ApJ...494..139J} used the inhomogeneous conical jet model proposed by \citet{1981ApJ...243..700K} to estimate the jet parameters for a sample of of AGNs, and then the jet power is derived \citep{2009MNRAS.396..984G}.  For M87, \citet{2015MNRAS.451..927Z} elaborately modeled the radio emission from the jets including the effect of radio-jet core shift, which provided an estimate of the jet power. It is in the range of the averaged jet power derived in previous works \citep{2002ApJ...579..560Y,2006MNRAS.372...21A,2012A&A...547A..56D}. 

The relativistic jets in AGNs are supposed to be most probably powered by taping energy from spinning BHs with corotating magnetic field, namely, the Blandford-Znajek (BZ) mechanism  \citep{1977MNRAS.179..433B}. The large-scale magnetic field is a key ingredient in this model, while its origin is still quite uncertain. One of the most promising scenarios is a weak external coherent field threading the gas being dragged inwards by the accretion disc  \citep[][]{1974Ap&SS..28...45B,1994MNRAS.267..235L,2011ApJ...737...94C,2023ApJ...944..182D}. It was suggested that a very weak external field with strength of tens $\mu\rm G$ can be significantly amplified in the disc, and it finally accelerates strong jets from the horizon of a rapidly spinning BH in the AGN \citep{2016ApJ...833...30C,2019MNRAS.485.1916C}.

The field strength increases with decreasing radius in the disc, and the field can be dynamical important in the inner region of the disc under certain circumstance, and the gas in the disc is magnetically arrested 
\citep{2003PASJ...55L..69N,2011ApJ...737...94C,2023ApJ...944..182D}. For stellar BHs, there is strong observational evidence of the formation of a magnetically arrested disc (MAD) in MAXI J1820$+$070 \citep{2023Sci...381..961Y}, while it is still lack of direct observational evidence of MAD in AGNs  \citep{2024ApJ...972...34L}. 
Even for one of the best observed radio galaxy, M87, it is still debating whether a MAD in it, though some investigations suggested evidence of a MAD in M87 by the comparison of the theoretical model calculations with the observed spectral data \citep[e.g.,][]{2022ApJ...924..124Y,2026A&A...705A.156S,2026A&A...708A.192G,2024SciA...10N3544Y}.

In this work, we calculate the magnetic field advection in the accretion disc (either a MAD or a standard normal evolution disc) in M87 on the assumption of a weak external field is dragged inwards by the gas in the disc \citep[e.g.,][]{1999ApJ...512..100L}. The jet power is then calculated with derived field strength at the BH horizon, which is confronted with the jet power estimated from observations. The accretion state in M87 is therefore revealed. We describe our model calculations in Section \ref{sec:model}. The results are summarized in Section \ref{sec_results}, and Section \ref{sec_discuss} contains a brief discussion of the results.

\section{Model} \label{sec:model}

The jets from the central region near the BH have been observed in M87, which are widely believed accelerated by the magnetic fields, most probably via the BZ mechanism.  Strong magnetic fields responsible for jet formation can only be maintained by the currents in the accretion flow surrounding the BH.  One of the most promising scenario is a weak external coherent field threading the circumnuclear gas being advected inwards in the accretion disc.  We assume that a weak external field at the outer radius of the ADAF is advected inwards by the ADAF. As the strength of the external field is still uncertain, the field strength of the interstellar medium (ISM) in some galaxies has been constrained by the observations, which is up to $\sim 10^{-3} \rm Gauss$ \citep{2014ApJ...786....5K}.  Thus, we adopt the external field strength as a model parameter. 

The structure of the accretion flow in M87 has been inferred by multi-waveband continuum spectral fitting together with the radio polarization data, with which the field advection in the ADAF can be calculated when the magnetic diffusivity is given. The  magnetic diffusivity for turbulent plasma is related to the viscosity parameter $\alpha$ by the Prandtl number.  The field is usually substantially enhanced in the inner edge of the ADAF due to its large radial velocity, and the gas may be arrested if the magnetic pressure is comparable with the gas pressure.  The field strength at the BH horizon is estimated based on the calculations of the field advection in the ADAF, and then the BZ power is available as a function of the BH spin parameter $a$. The derived BZ power can be constrained by the observations of the jets in M87 (see Section \ref{sec_pjet_m87}).

\subsection{Advection dominated accretion flow with magnetic outflows}  \label{sec_adaf}

When magnetic outflows carry away a fraction of the angular momentum from the ADAF,  its radial velocity becomes higher than that of a conventional ADAF \citep{2009MNRAS.400.1734L}.  
Thus, the radial velocity of the ADAF with a magnetic field that accelerates the outflow could be described as 
\begin{equation}
v_{R}^{\prime} \simeq v_{R} + v_{R,\rm m}.
\label{eq:speed}
\end{equation}
The first term is radial velocity of a conventional ADAF driven by turbulence \citep{1994MNRAS.267..235L},
\begin{equation}
v_{R} \simeq -\frac{\nu}{R},
\label{eq:speed2}
\end{equation}
where the viscosity $\nu=\alpha c_{\rm s} H$ ($c_{\rm s}$ is the sound speed of the gas, and $H$ is the half-thickness of the disc).
The second term is contributed by the outflows \citep{2016ApJ...817...71C},
\begin{equation}
v_{R,\rm m} = -\frac{2T_{\rm m}}{R \Sigma \Omega},
\end{equation}
where $\Omega$ is the angular velocity of the gas in the accretion flow, and the magnetic torque exerted by the outflows on the unit area of the ADAF is
\begin{equation}
T_{\rm m}=\frac{B_{z}B_{\phi}^{\rm s}}{2\pi}R.
\end{equation}
Here $B_{\phi}^{\rm s}$ is the azimuthal component of the large-scale magnetic field at the disc surface and $B_{z}$ is the vertical component of the field.  

The radial velocity of ADAF with magnetically driven outflows is then given by 
\begin{align}
v_{R}^{'}&=v_{R}+v_{R, \rm m}=-\alpha c_{\rm s}\frac{H}{R}-\frac{2T_{\rm m}}{\Sigma R \Omega} \notag \\ 
&=-\alpha c_{\rm s}\frac{H}{R}-\frac{B_{z}B_{\phi}^{\rm s}}{\pi \Sigma \Omega}=v_{R}\left(1+\frac{B_{z}B_{\phi}^{\rm s}}{\pi \Sigma \Omega} \frac{R}{\alpha c_{\rm s}H}\right). \label{eq:vr0}
\end{align} 
Equation (\ref{eq:vr0}) can be re-written as 
\begin{equation}
v_{R}^{\prime} = (1+f_{\rm m})v_{R}=-(1+f_{\rm m})\alpha c_{\rm s}\frac{H}{R},
\label{eq:VR}
\end{equation}
where the relative importance of the outflows on the radial velocity of the ADAF,
\begin{equation}
f_{\rm m}=\frac{4\xi _{\phi }}{\tilde{H}\beta \alpha f_{\Omega}}.
\label{eq:fm}
\end{equation}
The dimensionless quantities are defined as $\xi_{\phi}=-B_{\phi}^{\rm s}/B_{z}$, $\beta=P_{\rm gas}/{(B_{z}^2/8\pi)}$, $\tilde{H}=H/R$, 
and the ratio $\xi_{\phi}\lesssim 1$ is required by the stability criterion \citep[e.g.,][]{1999ApJ...512..100L}. For an ADAF with $\tilde{H}\sim 1$ and $f_{\Omega}\lesssim 1$, we can estimate the value of  $f_{\rm m}\sim4/\beta\alpha$.

The large-scale magnetic field threading the ADAF provides a radial  counterforce against gravity, which will alter the dynamics of the ADAF significantly if the field is sufficiently  strong \citep[][]{2003PASJ...55L..69N}.

The magnetic force exerted on the unit surface area of the disc is \citep{2003PASJ...55L..69N}
\begin{equation}
F_{\rm magnetic} \sim \frac{2B_{R}^{\rm s}B_{z}}{4\pi } \sim \frac{B_{z}^2}{2\pi},
\label{eq:mag}
\end{equation}
where $B_R^{\rm s}$ is the strength of the radial component of the field at the disc surface. In ADAF, the magnetic field lines are stretched by radial flow, and typically $B_R^s \sim B_z$ \citep{2003PASJ...55L..69N}.  The radial gravitational force exerted on per unit area of the disc is
\begin{equation}
F_{\rm gravity} \sim \frac{GM \Sigma }{R^{2} },
\label{eq:gra}
\end{equation}
while the centrifugal force of the rotating disc is
\begin{equation}
F_{\rm Centrifugal} \sim  \Sigma \Omega^2 R.
\label{eq:cen}
\end{equation}
Thus, the accreting gas may be magnetically arrested when the field is sufficiently strong, i.e., 
\begin{equation}
F_{\rm magnetic} \gtrsim  F_{\rm gravity}-F_{\rm Centrifugal}. 
\label{eq:mad_condi}
\end{equation}
In this case, the gas moves inwards through interchange instability, and the radial velocity $v_R$ of the gas is very small, which is at the order of  $-\epsilon v_{\rm {ff}}$,  $v_{\rm ff} \sim (GM/R)^{1/2}$  is the free-fall velocity, and $\epsilon \sim 0.001 - 0.01$ \citep{2003PASJ...55L..69N}.  Substituting Equations (\ref{eq:mag})-(\ref{eq:cen}) into Equation (\ref{eq:mad_condi}), we obtain the MAD condition for the disc, 
\begin{equation}
\frac{B_{z}^2}{2\pi}\gtrsim  \frac{B_{\rm mad}^2}{2\pi}=\frac{GM\Sigma }{R^{2} }-\Sigma \Omega^2 R=\frac{GM \Sigma }{R^{2}} (1-{f_{\Omega}}^2),
\label{eq:MAD} 
\end{equation}
where $f_{\Omega}={\Omega}/{\Omega_{K}}$. 
The surface density $\Sigma$ of the disc is given by 
\begin{equation}
\Sigma=-\frac{\dot{M}(R)}{2\pi Rv_{R}}=-\frac{\dot{M}(R)}{2 \pi R \epsilon v_{\rm {ff}}}. \label{eq:sigma}
\end{equation}
where $\dot{M}(R)$ is the accretion rate at $R$.  Substituting Equation (\ref{eq:sigma}) into (\ref{eq:MAD}), the MAD condition becomes
\begin{equation}
B_{\rm mad} \sim 1.4\times 10^{9}(1-{f_{\Omega}}^2)^{1/2}\epsilon^{-1/2}m^{-1/2}\dot{m}(r)^{1/2}r^{-5/4} \rm Gauss,
\label{eq:Bmad}
\end{equation}
where $m=M_{\rm BH}/M_{\odot}$, $\dot{m}=\dot{M}/{\dot{M}_{\rm Edd}}$, $\dot{M}_{\rm Edd}=L_{\rm Edd}/0.1c^2$ is Eddington accretion rates, and $r=Rc^2/GM_{\rm BH}$. 
The accretion flow would be magnetically arrested in the region where the magnetic field is stronger than $B_{\rm mad}$, and its radial velocity becomes substantially low, $v_R\sim \epsilon v_{\rm ff}$ \citep{2003PASJ...55L..69N}. The physics of the mass transfer in the MAD region is complicated, and we adopt 
\begin{equation}
v_{R}^{\prime}=\left\{\begin{array}{ll}
\left(1+f_{\rm m}\right) v_{R} =-\left(1+f_{\rm m}\right)\alpha c_{\rm s}\frac{H}{R}& \text { if } R>R_{\rm m} \\
-\epsilon  v_{\rm ff} & \text { if } R \leq R_{\rm m}
\end{array}\right.,
\label{eq:vel}
\end{equation}
for simplicity \citep{2023Sci...381..961Y}, which can describe the main feature of the radial velocity of a MAD.

\subsection{Formation of magnetic field in the ADAF} \label{sec_b_field}

In this work, we assume that an external vertical field is advected inwards by the ADAF, while the field will diffuse out at the same time, which is described by the magnetic diffusivity in a turbulent accretion flow. Thus, the advection/diffusion of the field in an ADAF with magnetic outflows is described by the induction equation, 
\begin{equation}
\frac{\partial}{\partial t}[R \psi(R, 0)]=-v_{R}^\prime \frac{\partial}{\partial R}[R \psi(R, 0)]-\frac{4 \pi \eta_{\rm m}}{c} \frac{R}{2 H} \int\limits_{-H}^{H} J_{\phi}\left(R, z_{\rm h}\right) d z_{\rm h},
\label{eq:advection}
\end{equation}
where $J_{\phi}(R,z_{\rm h})$ is the current density at $z = z_{\rm h}$ above or below the mid-plane of the disc, $\eta_{\rm m}$ is the magnetic diffusivity, $\psi(R,z)$ is the azimuthal component of the magnetic potential, and the vertical velocity $v_{z}$ of the flow is neglected \citep[see][for the details]{2011ApJ...737...94C}. Assuming the azimuthal current distribution in the vertical direction to be homogeneous, we have 
\begin{equation}
J_{\phi}(R,z_{\rm h})=\frac{J_{\phi}^{\rm s} (R)}{2H},
\end{equation}
where $J_{\phi}^{\rm s} (R)$ is the surface current density of the disc.

The magnetic field potential $\psi(R, z) = \psi_{\rm d}(R, z) + \psi_{\rm \infty }(R, z)$, where $\psi_{\rm d}(R, z)$ is contributed by the currents in the accretion flow, and $\psi_{\rm \infty }(R) = B_{\rm 0}R/2$ is the external imposed homogeneous vertical field, which can be regarded as being contributed by the currents at infinity \citep{1994MNRAS.267..235L}.The potential $\psi_{\rm d}$ is related to $J_{\phi}^{\rm s} (R)$ with
\begin{equation}
    \begin{aligned}
\psi_{\rm d}(R, z)=&\frac{1}{c} \int\limits_{R_{\text {in }}}^{R_{\text {out }}} R^{\prime} \rm d R^{\prime} \int\limits_{0}^{2 \pi} \cos \phi^{\prime} \rm d \phi^{\prime} \\
&\quad \times \int\limits_{-H}^{H} \frac{J_{\phi}\left(R^{\prime}, z_{\rm h}\right)}{\left[R^{\prime 2}+R^{2}+\left(z-z_{\rm h}\right)^{2}-2 R R^{\prime} \cos \phi^{\prime}\right]^{1 / 2}} \rm{~d} z_{\rm h} \\
=&\frac{1}{2 H c} \int\limits_{R_{\text {in }}}^{R_{\text {out }}} J_{\phi}^{\rm s}\left(R^{\prime}\right) R^{\prime} \rm d R^{\prime} \int\limits_{0}^{2 \pi} \cos \phi^{\prime} \rm d \phi^{\prime} \\
&\quad \times \int\limits_{-H}^{H} \frac{1}{\left[R^{\prime 2}+R^{2}+\left(z-z_{\rm h}\right)^{2}-2 R R^{\prime} \cos \phi^{\prime}\right]^{1 / 2}} \rm{~d} z_{\rm h}.
    \end{aligned}
\label{eq:psid}
\end{equation}{}

Differentiate Eq.(\ref{eq:psid}), we have \citep{2011ApJ...737...94C}
\begin{equation}
    \begin{aligned}
\frac{\partial }{\partial R} &[R\psi_{\rm d}(R, z)]=\frac{1}{2 H C} \int\limits_{R_{\text {in }}}^{R_{\text {out }}}J_{\phi}^{\rm s}\left(R^{\prime}\right) R^{\prime} \rm d R^{\prime} \int\limits_{0}^{2 \pi} \cos \phi^{\prime} \rm d \phi^{\prime} \\
&\quad \times \int\limits_{-H}^{H} \frac{R^{\prime 2} + \left(z-z_{\rm h}\right)^{2} - R R^{\prime} \cos \phi^{\prime}}{\left[R^{\prime 2}+R^{2}+\left(z-z_{\rm h}\right)^{2}-2 R R^{\prime} \cos \phi^{\prime}\right]^{3 / 2}} \rm{~d} z_{\rm h}.
    \end{aligned} \label{eq:dpsidr}
\end{equation}
For steady case that $\partial / \partial t =0$, Eq.(\ref{eq:advection}) becomes
\begin{equation}
-{\frac{\partial }{\partial R}} [R\psi_{\rm d}(R, 0)]-{\frac{2\pi}{c}}{\frac{\alpha c_{\rm s}(R)R}{{v}_{R}(R)}}P_{\rm m}J_{\phi}^{\rm s}(R)=B_{0}R,
\label{eq:steady case}
\end{equation}
where the magnetic Prandtl number $P_{\rm m}\equiv \eta_{\rm m}/\nu$. 

 {The magnetic diffusivity $\eta_{\rm m }$ is related to the Ohmic dissipation in the plasma. 
\citet{1979cmft.book.....P} argued that the viscosity $\nu\sim \eta_{\rm m} \sim l v_{\rm t}$ ($l$ is the largest eddy size and $v_{\rm t}$ is turnover velocity), and then $P_{\rm m}\equiv \eta_{\rm m}/\nu\sim 1$, is expected in isotropic turbulence. This issue was explored by different authors using numerical simulations \citep[e.g.,][]{2003A&A...411..321Y, 2009ApJ...697.1901G,2009A&A...504..309L,2009A&A...507...19F}, which suggest that the magnetic Prandtl number should be in the range of $\sim 0.2-1$. Thus, the magnetic diffusivity $\eta_{\rm m}$ in an astrophysical accretion disc is parametrized in terms of the MRI turbulent viscosity through the magnetic Prandtl number $P_{\rm m}\equiv \eta_{\rm m}/\nu$. }

For an ADAF with magnetic outflows, the radial velocity is give by Equation (\ref{eq:VR}). Substituting Equation (\ref{eq:dpsidr}) into (\ref{eq:steady case}), we have
\begin{equation}
    \begin{aligned}
&B_{0}R=-\frac{1}{2Hc}\int\limits_{R_{\text {in }}}^{R_{\text {out }}} \int\limits_{0}^{2\pi} J_{\phi}^{\rm s}(R^{\prime})cos\phi^{\prime}\rm d\phi^{\prime} R^{\prime} \rm d R^{\prime}\\
&\times \int\limits_{-H}^{H}\frac {{R^{\prime 2}}+{z_{\rm h}}^2-RR^{\prime}cos\phi^{\prime}}{[{{R^{\prime 2}}+R^2+{z_{\rm h}}^2-2RR^{\prime}cos\phi^{\prime}]}^{3/2}}d z_{\rm h}-\frac{2\pi}{c}\frac{\alpha c_{\rm s}(R)R}{{v}_{R}(R)}P_{\rm m}J_{\phi}^{\rm s}(R),
    \end{aligned}
\label{eq:integral-differential}
\end{equation}
which can reduce to a set of linear algebraic equations,
\begin{equation}
-\sum_{j=1}^{n} P_{ij}J_{\rm \phi}^{\rm s}(R_{j})\Delta R_{j}- \frac{2\pi}{c}\frac{\alpha c_{s}(R_{i})R_{i} }{v_{R}^{\prime}(R_{i})}P_{\rm m} J_{\rm \phi}^{\rm s}(R_{i})=B_{0}R_{i},
\label{linear algebraic equations}
\end{equation}
where $J_{\rm \phi}^{\rm s}(R_{j})$ is the surface current density of the ring at radius $R_{j}$ in the accretion flow, $\Delta R_{j}$ is the width of the ring, and 
\begin{equation}
P_{ij}=\frac{R_{j} }{2H_{j} c } \int\limits_{0}^{2\pi}cos\phi^{'} d\phi^{'} \int\limits_{-H_{j}}^{H_{j}}\frac{R_{j}^2+z_{\rm h}^2-R_{i}R_{j}cos\phi^{'}}{[R_{i}^2+R_{j}^2+z_{\rm h}^2-2R_{i}R_{j}cos\phi^{'}]^{3/2}} dz_{\rm h}. 
\end{equation}

Given the ADAF structure and the specified magnetic Prandtl number $P_{\rm m}$, the surface current density $J_{\rm \phi}^{\rm s}(R_{i})$ can be calculated by solving a set of linear algebraic equations \citep{1994MNRAS.267..235L}.

The configuration of the large-scale magnetic fields can be calculated with the derived potential $\psi(R,z)$,
\begin{equation}
B_{R}(R,z)=-\frac{\partial }{\partial z} \psi(R.z),       
\end{equation}
and
\begin{equation}
B_{z}(R,z)=\frac{1}{R}\frac{\partial }{\partial R} [R\psi(R.z)].       
\end{equation}

\subsection{Blandford-Znajek jet power} \label{sec_p_bz}

 {The BZ power is determined by two key quantities: the black hole spin $a$ and the large-scale poloidal magnetic flux $\Phi_{\rm BH}$ threading the horizon \citep{2015ASSL..414...45T},}
\begin{equation}
P_{\rm BZ}=\frac{c}{96 \pi^2 R_{\rm g}^2}\Phi_{\rm BH}^2 \omega_{\rm H}^2 ,
\end{equation}
where $\Phi_{\rm BH}=2\pi B_{\rm H}R_{\rm H}^2$ is the magnetic flux,  the normalized angular frequency at the event horizon of the BH is 
\begin{equation}
\omega_{\rm H}\equiv \frac{\Omega_{\rm H}}{\Omega_{\rm H,\rm max}}=\frac{a}{1+(1-a^2)^{1/2}},
\end{equation}
and the radius of the BH event horizon is 
\begin{equation}
R_{\rm H}=(1+(1-a^2)^{1/2})R_{\rm g}=(1+(1-a^2)^{1/2}{\frac {GM_{BH}}{c^2}}.
\end{equation}
The BH spin parameter is
\begin{equation}
a\equiv J_{\rm BH}/J_{\rm max}\equiv J_{\rm BH}/M_{\rm BH}r_{\rm g}c. 
\end{equation}
Thus, the jet power can be expressed as
\begin{equation}
P_{\rm BZ}=\frac{1}{24}{B_{\rm H}}^{2}{R_{\rm H}}^{2}ca^2.
\label{eq:jetpower}
\end{equation}

Our present calculations of the field advection in the ADAF are carried out in the Newtonian frame, in which the field strength at the innermost stable circular orbit (ISCO) is derived when the values of the model parameters are specified. The radius of the ISCO for a BH is 
\begin{equation}
r_{\rm isco}=R_{\rm isco}/R_{\rm g}=3+Z_{2}-{[(3-Z_{1})(3+Z_{1}+2Z_{2})]}^{1/2}. 
\label{ISCO}
\end{equation}
where $Z_{1}\equiv 1+(1-a^{2})^{1/3}[(1+a)^{1/3}+(1-a)^{1/3}]$, and $Z_{2}\equiv (3a^{2}+{Z_{1}}^{2})^{1/2}$ \citep[][]{1972ApJ...178..347B}.

A poloidal field is assumed to thread the horizon in the BZ power formula (Eq.~\ref{eq:jetpower}), which is twisted by the rotating BH to generate a toroidal component, and it is becoming toroidally dominated beyond the Alfv\'{e}n surface. The detailed physical processes of the BZ mechanism has been extensively studied with GRMHD simulations \citep[see][for a review]{2015ASSL..414...45T}. In this work, the advection of poloidal field in the accretion flow is calculated, and then the BZ power is derived with Equation (\ref{eq:jetpower}) to avoid the complexity of the jet acceleration near the BH horizon. 

It is worth noting that our model is formulated in a Newtonian framework, which is unable to describe the dynamics of the accretion flow in the plunge region between the ISCO and the BH horizon. The full general relativistic (GR) calculations are beyond the scope of this work. 
In this work, we only calculate the field advection in the accretion flow from the outer radius to ISCO, in which the transition of a normal ADAF (SANE) to a MAD near the ISCO is properly considered. The field will be further enhanced in the plunge region to the BH horizon, which is well studied  
with the general relativistic magnetohydrodynamics (GRMHD) simulations. It is found that the ratio of the field strength at the $R_{\rm H}$ to $R_{\rm isco}$ is $\sim 3-6$,  when the BH spin parameter increases from $a=0$ to $0.94$ \citep{2004ApJ...611..977M}. For a rapidly rotating BH in M87, we adopt $f_{\rm B}\sim 6$ ($B_{\rm H}=f_{\rm B} B_{\rm isco}$) in our calculations.

\subsection{Accretion Rate of the ADAF} \label{sec_mdot}

The Bondi accretion rate and radius can be estimated with 
\begin{equation}
\dot {M}_{\rm Bondi}=4\pi\lambda(GM_{\rm BH})^2 c_{\rm s}^{-3}\rho, 
\label{Mdot}
\end{equation}
at the Bondi radius, 
\begin{equation}
R_{\rm A}=2GM_{\rm BH}/c_{\rm s}^{2}, 
\label{RA}
\end{equation}
when the properties of the circumnuclear gas, i.e., the density and temperature, are known \citep{1952MNRAS.112..195B}, which can be derived from the extended soft X-ray emission for nearby galaxies \citep[][]{2006MNRAS.372...21A}. For M87, $\dot{M}_{\rm Bondi}\simeq 0.12 M_\odot~{\rm yr}^{-1}$ and $R_{\rm A} \simeq 8.37\times 10^5~R_{\rm g}$ are derived with the high-resolution X-ray observations, in which $M_{\rm BH}=6.5\times 10^9 M_\odot$ is adopted. If the gas at $R_{\rm A}$ has a certain amount of angular momentum, it will falls inwards almost radially till the circularization radius, which depends on the specific angular momentum of the gas at $R_{\rm A}$ \citep{2008MNRAS.383..458C,2016ApJ...833...30C}. 

The angular momentum depends sensitively on its origin. For Sgr A$^*$, its circumnuclear hot gas most probably originates from the strong winds produced by massive stars in the central region, therefore the average angular momentum of the gas should be rather small due to chaotic motions of the stars \citep{1991ApJ...382L..19K,1992ApJ...387L..25M,1997A&A...325..700N}. 
The Bondi accretion model can describe the gas falling from the $R_{\rm A}$ fairly well \citep{2004ApJ...613..322Q}. For M87, at the centre of the galactic cluster, besides the origin of stellar winds, the contribution of the galactic inflow to the circumnuclear extended gas may not be neglected, which makes the specific angular momentum of the gas at the Bondi radius be quite uncertain. However, the constraint on the angular momentum of the gas at Bondi radius is derived for elliptical galaxy NGC 4278, which may provide useful clues. The accretion rate of the BH in NGC 4278 is estimated by spectral fitting of its nuclear X-ray emission with ADAF model, which is close to the Bondi accretion rate derived with the X-ray observations on the extended hot gas \citep{2012ApJ...758...94P}. It is natural to conclude that the angular momentum of the gas at the Bondi radius must be very small. Otherwise, the real accretion rate at the Bondi radius will be lower than the Bondi accretion rate due to the centrifugal force caused by gas rotation against the gravity \citep{2014ARA&A..52..529Y}. It is quite unlikely that the accretion rate of the ADAF is higher than that at the Bondi radius. Unfortunately, such approach is inapplicable for M87, because the BH accretion rate is significantly lower than its Bondi accretion rate. The specific angular momentum of the gas may vary for individual galaxies, but should not be very large if the gas distributions are similar in elliptical galaxies. 

It is unlikely to measure the specific angular momentum of the gas at the Bondi radius in M87 directly, we assume that the circularization radius is approximately one-tenth of the accretion radius as adopted in some previous works \citep[e.g., see][and the references therein]{2014ARA&A..52..529Y}. We adopt the circularization radius as the outer radius of the ADAF, $R_{\rm out} \simeq 0.1R_{\rm A} \simeq 8.37\times 10^4~R_{\rm g}$. We also explore how the results are affected by the outer radius of the accretion flow, which are summarized in Sect. \ref{sec_results}.

ADAFs are hot, which may possess strong outflows \citep[][]{1995ApJ...452..710N,1999MNRAS.303L...1B}, while their detailed properties are quite uncertain due to complicated physical processes in the ADAFs. As most gravitational energy of the accreting gas is released in the inner region of the ADAF, the accretion rate near the BH can be constrained by the multi-wavelength spectral energy distribution (SED). For M87, the accretion rate at $R = 10\,R_{\rm g}$ is derived with the multi-wavelength spectral model fits by incorporating Faraday rotation measurement (RM) \citep{2014ApJ...783L..33K},  which is in the range between $\sim 5.8\times10^{-4}-8.5 \times10^{-3}~ M_\odot\,\rm yr^{-1}$, due to the degeneracy of the parameters of the ADAF \citep{2016ApJ...830....6F}.

We adopt a power-law radially dependent accretion rate,  
\begin{equation}
\dot{M}(R)=\left\{\begin{array}{ll}
\dot{M}_{\rm out}\left(R/R_{\rm out}\right)^{p_{\rm w}}, 
& \text { if } R > R_{\rm m}, \\[8pt]
\dot{M}(R_{\rm m}), 
& \text { if } R \leq R_{\rm m} \;\; (p_{\rm w}=0).
\end{array}\right.
\label{eq:pw}
\end{equation}
to describe the mass loss in outflows, where the index $p_{\rm w}$ is available with the accretion rate at $R = 10\,R_{\rm g}$ and the Bondi rate \citep{1999MNRAS.303L...1B,2009MNRAS.400.1734L}. 
Accretion in the MAD region proceeds via interchange instability rather than MRI-driven turbulence, whether the interchange instability drives mass outflows remains unclear. In this work, we assume 
$\dot{M}(R \leq R_{\rm m})=\dot{M}(R_{\rm m})$, corresponding to no mass outflow within the MAD zone, in the region of  $R < R_{\rm m}$. Our calculations show that the MAD region is very narrow, only a few gravitational radii extending out from the ISCO, the field advection is therefore hardly affected even if outflows are present in the MAD region (see Section \ref{sec_results} for the details).

In this work, we adopt the accretion rates in the inner region of the ADAF in M87 derived with the model calculations proposed by \citet{2016ApJ...830....6F}, which are given as follows: 
\begin{itemize}
    \item \textbf{Model A} ($\delta=0.3$, $\beta_{\rm b} = 1$): $\dot{M}(10\,R_{\rm g})=5.8\times10^{-4}\,M_\odot\,\rm yr^{-1}$, corresponding to $p_{\rm w}=0.588$;
    \item \textbf{Model B} ($\delta=0.3$, $\beta_{\rm b} = 9$): $\dot{M}(10\,R_{\rm g})= 9 \times10^{-4}\,M_\odot\,\rm yr^{-1}$, corresponding to $p_{\rm w}=0.539$;
    \item \textbf{Model C} ($\delta=0.2$, $\beta_{\rm b} = 9$): $\dot{M}(10\,R_{\rm g})= 1.8 \times10^{-3}\,M_\odot\,\rm yr^{-1}$, corresponding to $p_{\rm w}=0.463$;
    \item \textbf{Model D} ($\delta=0.1$, $\beta_{\rm b} = 9$): $\dot{M}(10\,R_{\rm g})= 8.5 \times10^{-3}\,M_\odot\,\rm yr^{-1}$, corresponding to $p_{\rm w}=0.291$,
\end{itemize}
where different values of the electron heating fraction $\delta$ and the ratio of gas to magnetic pressure $\beta_{\rm b}$ for chaotic magnetic field in the accretion flow are adopted. We label these models as "Model A, B, C, D" hereafter.

\subsection{Structure of the ADAF in M87} \label{sec:stru_adaf_m87}

With the accretion rate as a function of radius derived from the observations of M87 (see Section \ref{sec_mdot}), the gas pressure of the ADAF with 
magnetic outflows can be expressed as 
\begin{equation}
    \begin{aligned}
&P_{\rm gas} \simeq 7.32\times10^{-2}\\
&\times \alpha^{-1} (1+f_{\rm m})^{-1} \dot{M}(r) \left(\frac{M_{\rm BH}}{M_{\odot}}\right)^{-2} {\left(\frac{R}{R_{\rm g}}\right)}^{-5/2} {\left(\frac{H}{R}\right)}^{-1} \rm g~  cm^{-1}~ s^{-2}.
\label{eq:gas}
    \end{aligned}
\end{equation}

Substituting the relation $\beta=8\pi P_{\rm gas}/{B_{z}^2}$ into Equations (\ref{eq:MAD}), we re-write the MAD condition as 
\begin{equation}
\beta \sim \frac{2\tilde{H}}{1-{f_{\Omega}}^2}. \label{eq:MAD_2}
\end{equation}

\begin{figure*}
    \centering
    \begin{subfigure}[b]{0.49\textwidth}
        \centering
        \includegraphics[width=\linewidth]{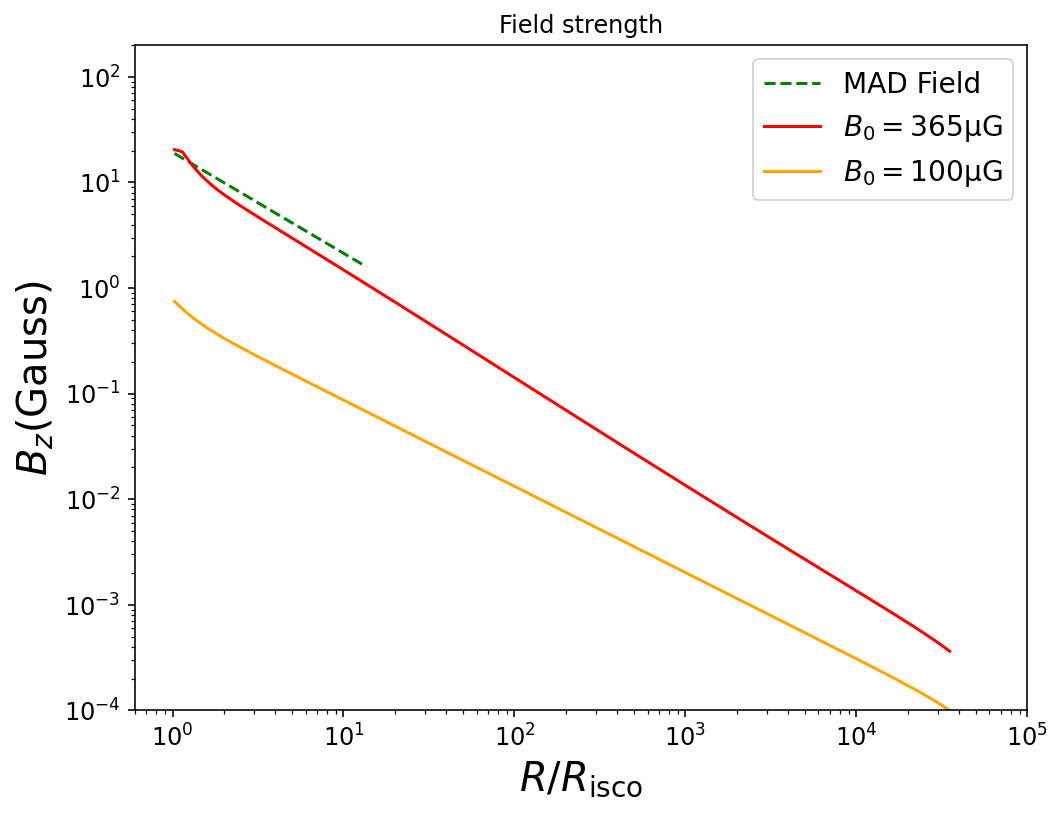}
        \caption{Model A}
    \end{subfigure}
    \hfill
    \begin{subfigure}[b]{0.49\textwidth}
        \centering
        \includegraphics[width=\linewidth]{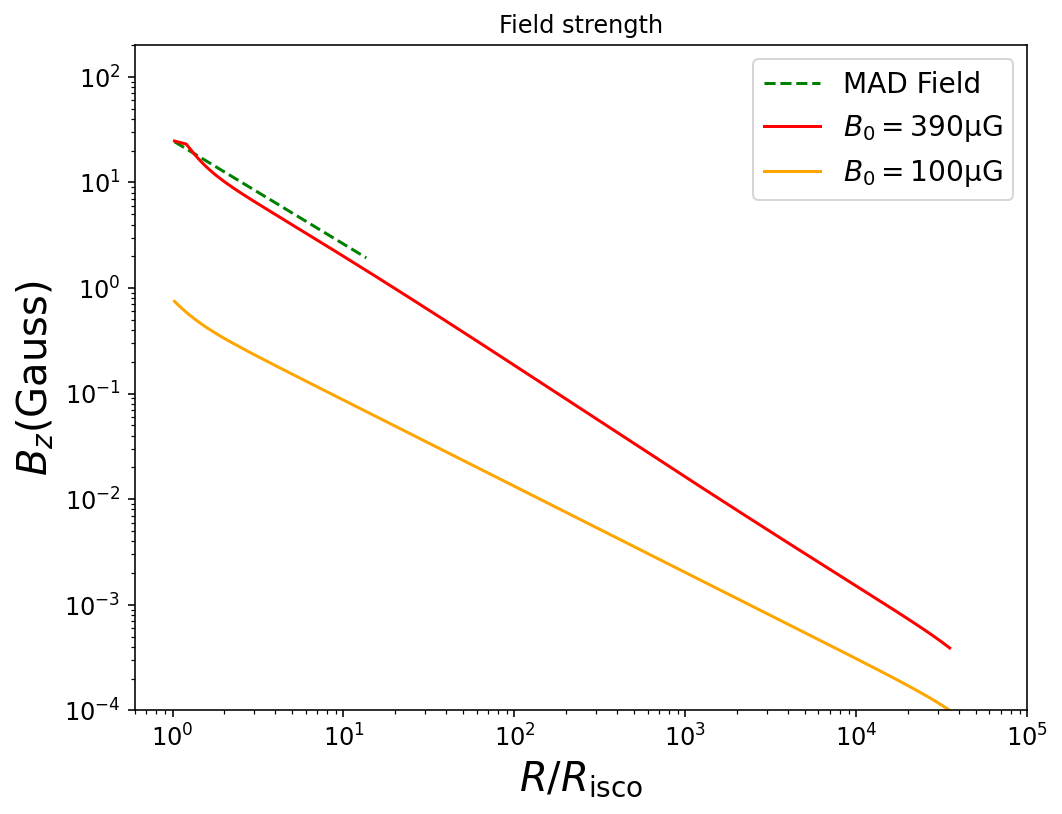}
        \caption{Model B}
    \end{subfigure}
    
    \vspace{5pt}
    
    \begin{subfigure}[b]{0.49\textwidth}
        \centering
        \includegraphics[width=\linewidth]{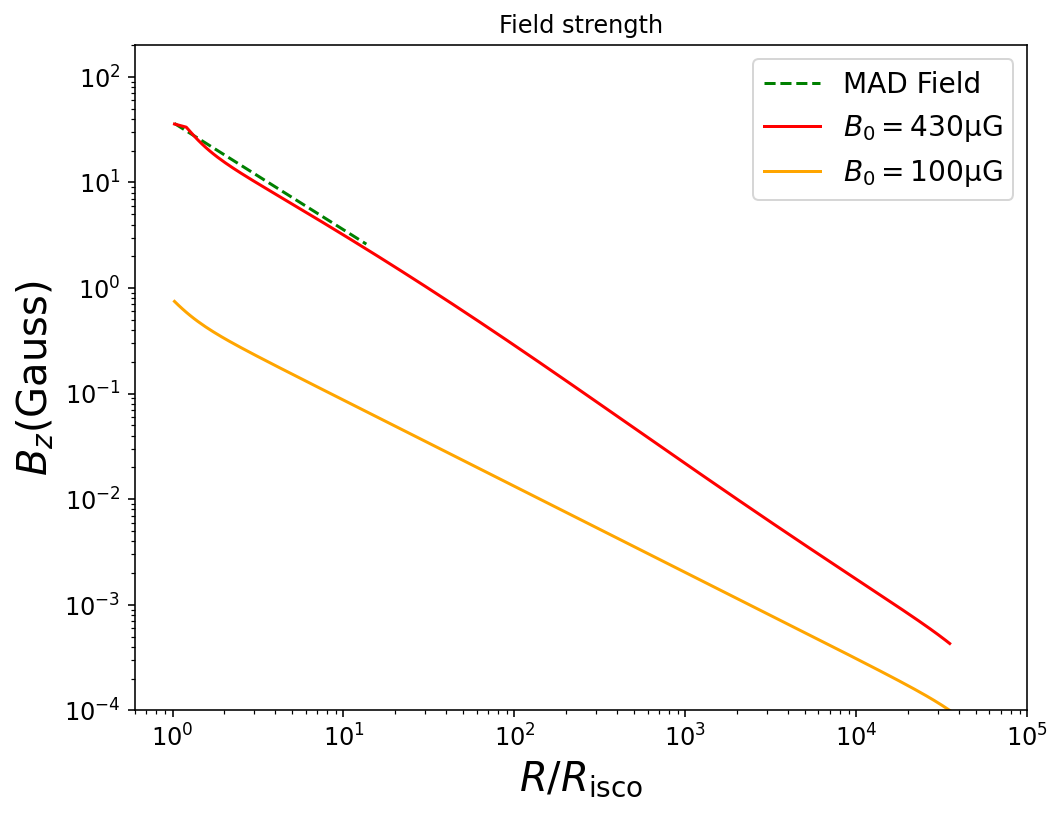}
        \caption{Model C}
    \end{subfigure}
    \hfill
    \begin{subfigure}[b]{0.49\textwidth}
        \centering
        \includegraphics[width=\linewidth]{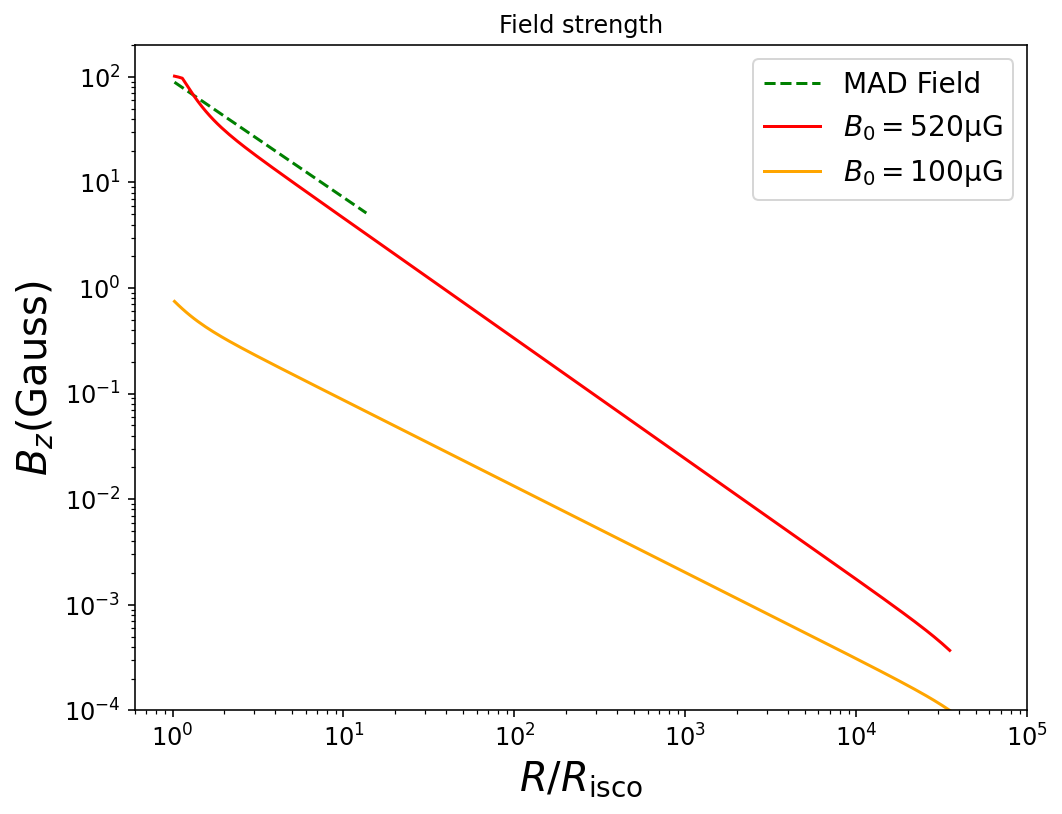}
        \caption{Model D}
    \end{subfigure}
    
    \caption{The vertical magnetic field strength in the accretion flow varies with radius. Model A, B, C and D respectively.  
    The magnetic Prandtl number $P_{\rm m} = 0.5$ is adopted in the calculations. The lines with different colours correspond to different values of external magnetic field strength $B_{0}$. The green dashed lines represent the the critical threshold for the magnetically arrested disc (MAD) state.}
    \label{fig:BZ}
\end{figure*}

\begin{figure*}
    \centering
    \begin{subfigure}[b]{0.49\textwidth}
        \centering
        \includegraphics[width=\linewidth]{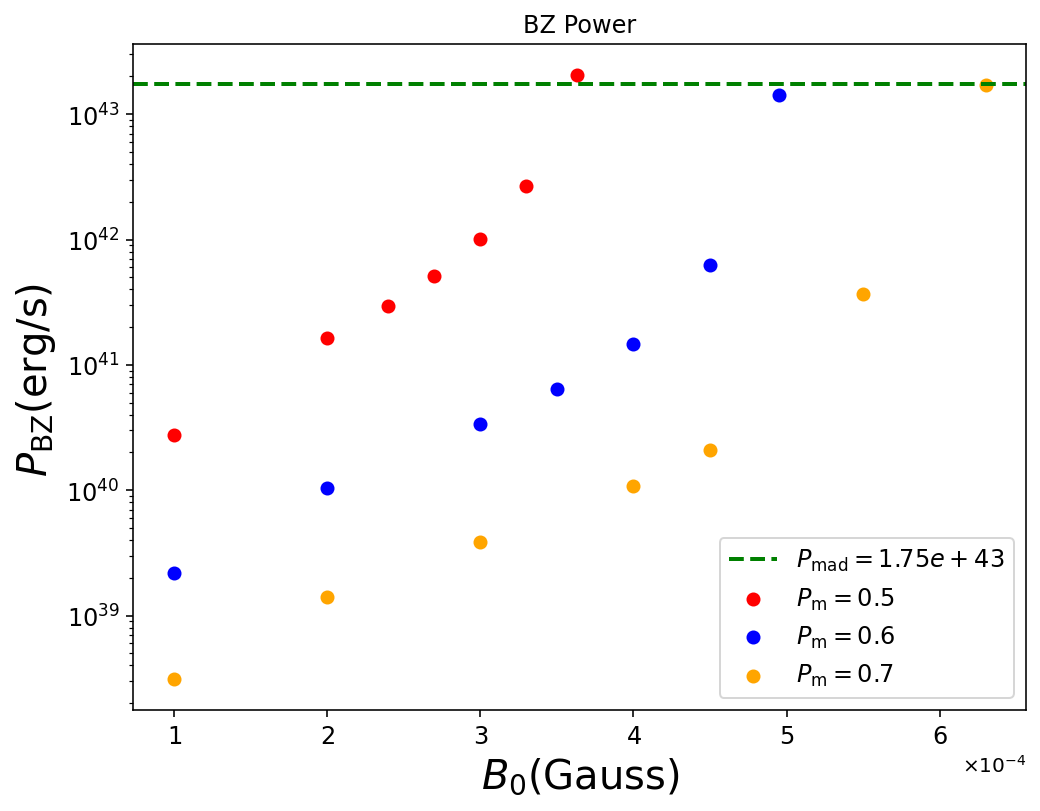}
        \caption{Model A}
    \end{subfigure}
    \hfill
    \begin{subfigure}[b]{0.49\textwidth}
        \centering
        \includegraphics[width=\linewidth]{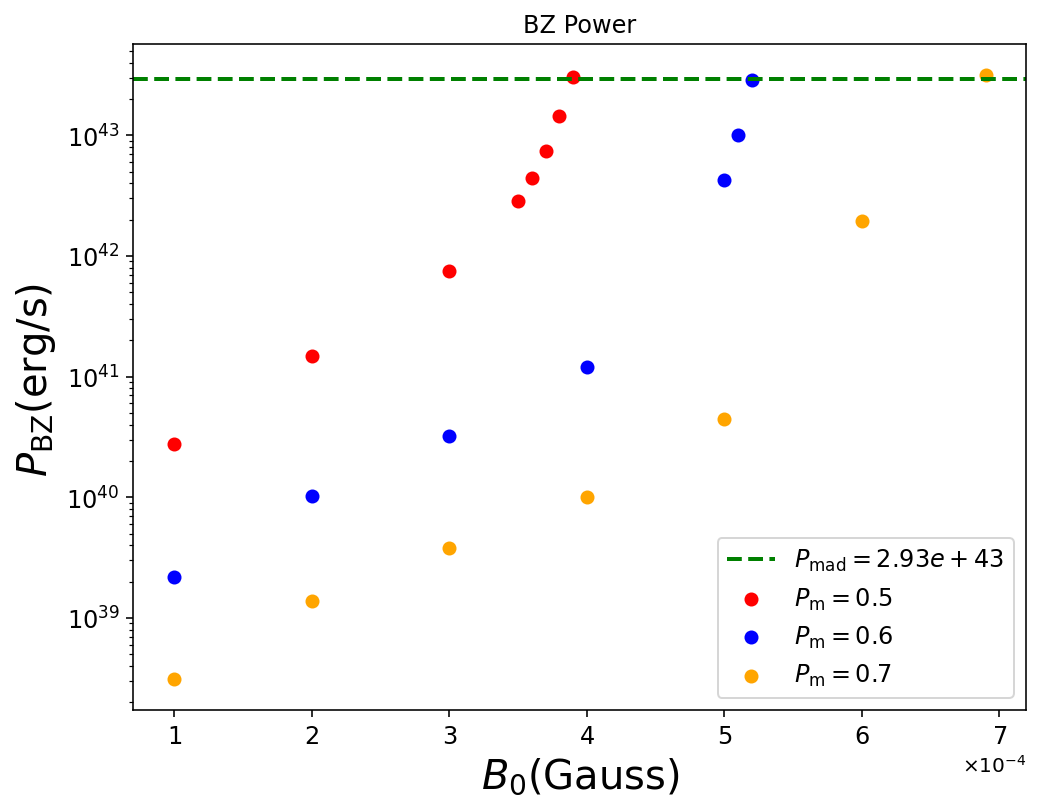}
        \caption{Model B}
    \end{subfigure}
    
    \vspace{5pt}
    
    \begin{subfigure}[b]{0.49\textwidth}
        \centering
        \includegraphics[width=\linewidth]{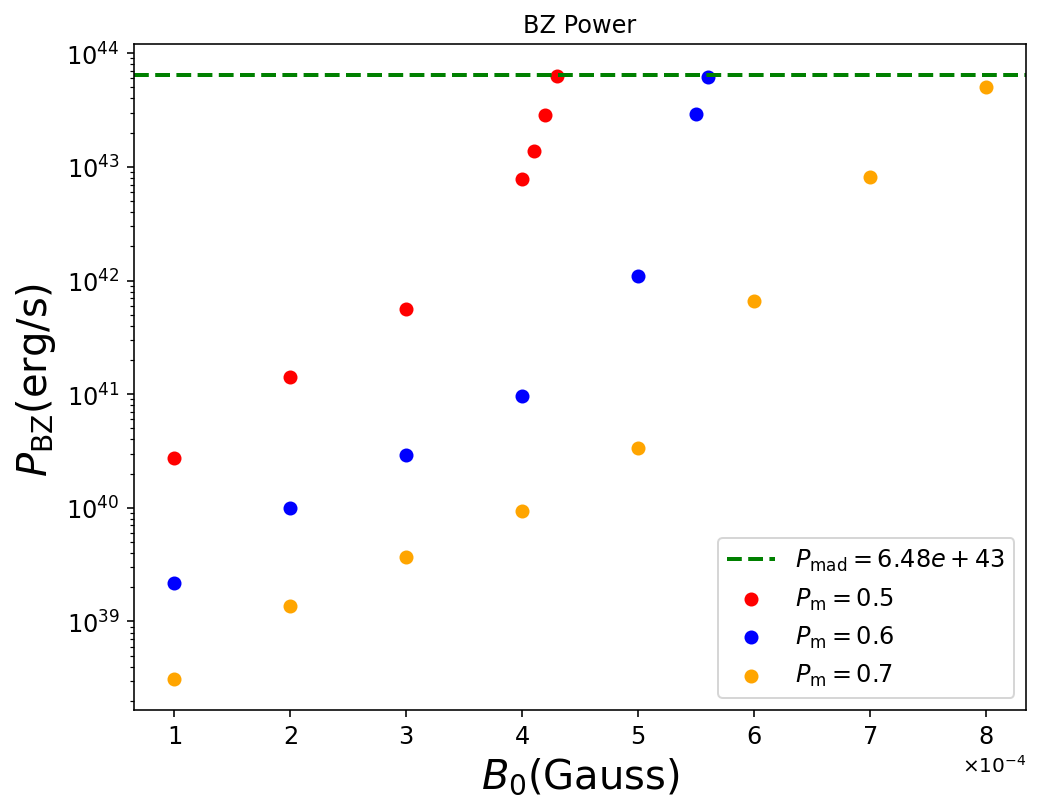}
        \caption{Model C}
    \end{subfigure}
    \hfill
    \begin{subfigure}[b]{0.49\textwidth}
        \centering
        \includegraphics[width=\linewidth]{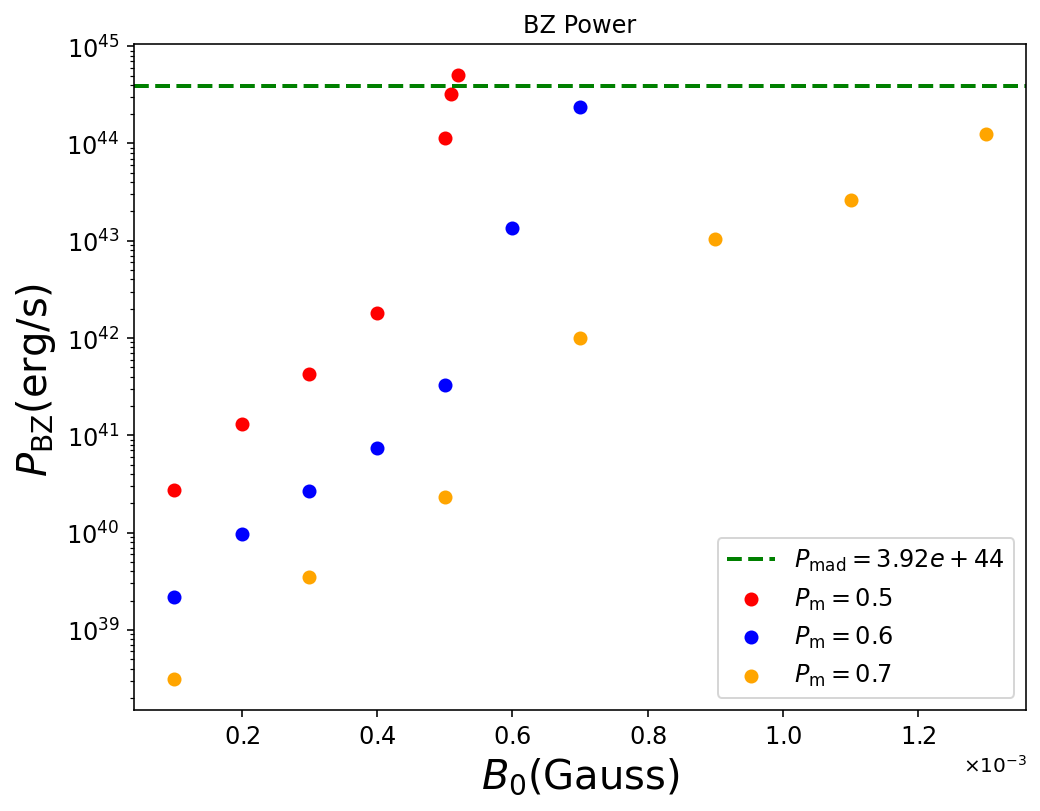}
        \caption{Model D}
    \end{subfigure}
    
    \caption{The Bransford-Zaneck (BZ) jet powers vary with the model parameters. 
    The lines with different colours correspond to different values of magnetic Prandtl number $P_{\rm m}$. The green dashed lines represent the critical BZ jet power for MAD state.}
    \label{fig:P_Bz}
\end{figure*}

\begin{figure} 
\centering 
\includegraphics[width=\linewidth]{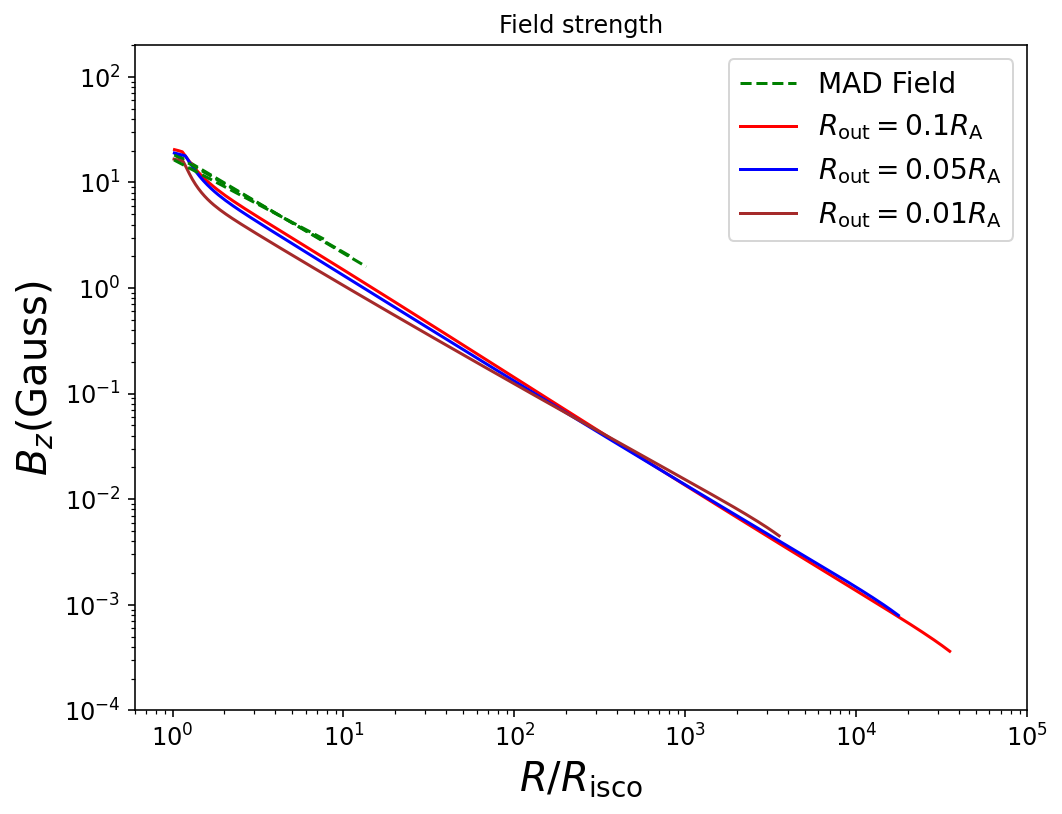}
\caption{The vertical magnetic field strength in the accretion flow varies with different circularization radius. 
The magnetic Prandtl number $P_{\rm m} = 0.5$ is adopted in the calculations. The red line, yellow line and blue line respectively represent $R_{\rm out}=0.1R_{A}$, $R_{\rm out}=0.05R_{A}$ and $R_{\rm out}=0.01R_{A}$. The green dashed lines represent the the critical threshold for the magnetically arrested disc (MAD) state.}
\label{fig:diff r_out}
\end{figure}

\section{Jet power of M87}  \label{sec_pjet_m87}

It is a challenging task to accurately measure the  jet power of M87.  There are cavities observed in the X-ray wavebands  in galaxy clusters, which are supposed to be formed by the energy injection by the jets.   The total energy required to produce a cavity is estimated as 
\begin{equation}
E_{\rm cav}=\frac{\gamma}{\gamma-1}PV,
\end{equation}
where $P$ is the pressure of the gas surrounding the bubble, $V$ is the volume of the bubble and $\gamma=4/3$ is the ratio of specific heats of the gas inside the cavity  \citep{2006ApJ...652..216R}. The temperature and density of the hot gas in the cavity can be inferred from the X-ray observations, and the volume $V$ is estimated from the high-resolution image of the cavity. Thus, the total output energy of the jets is derived. In order to derived the jet power, one needs to estimate the duration of the jet activity.   

There are three methods to estimate the formation time-scale of a cavity \citep{2004ApJ...607..800B}. These estimated values usually differ by a factor of $2-4$. The first is to assume that the bubble moves outward at the sound speed, and the age of the cavity is $t_{\rm age} = R/c_{\rm s}$, where $R$ is the distance from the edge of the cavity to the BH. The estimated jet power 
\begin{equation}
P_{\rm jet}\sim E_{\rm cav}/t_{\rm age}\sim3.5\times10^{43}\rm erg/s,
\end{equation}
for M87 \citep{2006MNRAS.372...21A}. The second estimate is based on the assumption that the cavity is a buoyant bubble rising at its buoyancy velocity, so its age is estimated as $t_{\rm age} = R/v_{\rm b}$. The buoyancy age could be the upper limit of the cavity's age. The estimated jet power is \citep{2006ApJ...652..216R}
\begin{equation}
P_{\rm jet}\sim E_{\rm cav}/t_{\rm age}\sim6_{-0.9}^{+4.2}\times10^{42} \rm erg~s^{-1}.
\end{equation}
In the third method, the time required for the matter to refill the cavity as it moves outwards is adopted as the cavity age. For non-relativistic plasmas (most of the bubble content is thermal, 
$\Gamma = 5/3$), the age of the halo is estimated as $\sim 250~(P_{\rm j}/10^{44})^{-1}~\rm Myr$, while $\sim 400(P_{\rm j}/10^{44})^{-1} ~\rm Myr$ for the bubble dominated by relativistic particles and magnetic fields ($\Gamma = 4/3$). Based on the averaged age of the halo $t \simeq 40\rm Myr$, the jet power is estimated as
\begin{equation}
P_{\rm jet}\sim 6\times10^{44} \rm erg~s^{-1}(\Gamma=5/3),
\end{equation}
or
\begin{equation}
P_{\rm jet}\sim 10\times10^{44}\rm erg~s^{-1}(\Gamma=4/3),
\end{equation}
respectively \citep{2012A&A...547A..56D}. We note that the estimates of the jet power are quite uncertain, and the derived jet power is the value averaged over a long period of time, which seems inadequate for our present work.

The radio core shift of relativistic jets is essentially an observational phenomenon caused by the synchrotron self-absorption (SSA) effect. For radio emission at different frequencies, the peak emission occurs at different distances along the jet axis, which ultimately manifests as the characteristic that the apparent position of the radio core varies inversely with the observing frequency \citep{1979ApJ...232...34B}. 

Since the emission from the radio core region originates from the innermost part of the jet adjacent to the black hole, corresponding to the zone with thousands of gravitational radii outside the black hole’s event horizon. The plasma and radiative properties of this region directly trace the ongoing jet emission from the black hole-accretion disc system, with a radiative timescale of merely days to years, which is fully consistent with the jet’s short-timescale activities. For this reason, it calibrates the instantaneous jet power and enables real-time tracking of the dynamic evolution of jet activity \citep{2014Natur.510..126Z}.

\citet{2015MNRAS.451..927Z} derived a novel method for magnetic field measurement based on the classical steady-state relativistic jet model established by Blandford and Königl \citep{1979ApJ...232...34B}. With the angular displacement of the core shifts obtained from VLBI, this method enables direct and accurate constraint on the magnetic field strength of the jet at a given distance without relying on the equipartition assumption. 

Combined with the physical correlation between the jet opening angle and the magnetization parameter, \citet{2015MNRAS.451..927Z} calculated the poloidal magnetic flux of the jet through the transverse-averaged toroidal magnetic field. They anchored the quantitative relationship between the magnetic Poynting power and the kinetic power of jet particles, and finally derived the instant jet power of M87 is approximately $3.2 \times 10^{43} \rm erg/s$ \citep[see][for the details]{2015MNRAS.451..927Z}.


\section{Results} \label{sec_results}

In this work, we adopt $M_{\rm BH}=6.5\times10^9 M_{\odot}$, the BH spin parameter $a = 0.9$ for M87, the $\alpha$-viscosity parameter $\alpha = 0.3$ \citep{2016ApJ...830....6F}, $f_\Omega=0.9$ and a typical value of $H/R = 0.5$ in the calculations of the magnetic field strength in the ADAF. 
We adopt $P_{\rm m} = 0.5$ in most of our calculations, but also explore how the results will be affected by the value of $P_{\rm m}$.   

When the external field strength $B_0$ at the outer radius $R_{\rm out}$ of the ADAF is specified, the strength of the field dragged inwards by the ADAF can be calculated if its radial velocity distribution is known (see Section \ref{sec_b_field}).   The key point in the calculations of the field advection is the radial velocity of the ADAF,  which is also affected by the magnetic field. It means that the field advection is  strongly coupled with the structure of the ADAF.  

The accretion rate as a function of radius is derived with the method described in Section \ref{sec_mdot}.  We find that the ratio of  magnetic pressure to gas pressure increases in the inner region of the ADAF, where the magnetic force exerted on the gas in the ADAF becomes dynamical important under certain circumstance. In this case, the gas in the accretion flow may be choked when the field is sufficiently strong to meet the criterion of MAD (see Section \ref{sec_adaf}).  In this region, the radial velocity of the gas is very small compared to a normal ADAF, which will surely affect the field advection.  

In this work, we first calculate the magnetic field configuration/strength based on the structure of the ADAF without magnetic field. With the derived magnetic field, we re-calculate the structure of the ADAF with magnetic outflows as described in Section \ref{sec_adaf}.  Repeat the calculation of the field advection with this newly obtained ADAF solution with outflows, the field strength/configuration is derived.  We find that the final solution is available usually after several  iterations. With derived magnetic field strength at the BH horizon,  we use Eq.(\ref{eq:jetpower}) to calculate the corresponding Bransford-Zaneck (BZ) jet power, which is then compared with the 
observed jet power (see Section \ref{sec_pjet_m87}).

We plot the magnetic field strengths varying with radius of the ADAFs in Figure \ref{fig:BZ} for different sets of model parameters. It is found that the gas of the ADAF becomes magnetically arrested in the inner region near the BH, provided the external field strength $B_0$ is sufficiently large.  Besides the external field strength $B_0$,  the results also depend on some other model parameters. The parameter space of our model calculations in order to reproduce the observed jet power is depicted in Figure \ref{fig:P_Bz}.  

To examine the influence of the outer boundary truncation radius on the magnetic field configuration, we carry out calculations by setting the outer boundary to $R_{\rm out} = 0.01R_{\rm A}$, and $0.05R_{\rm A}$, for comparison with Model A. In order to reproduce the same jet power, we find that $B_{0}=4.5{\rm mG}$ ($R_{\rm out} = 0.01R_{\rm A}$), and $B_{0}=795\rm \mu G$ ($R_{\rm out} = 0.05R_{\rm A}$) are required, while $B_{0}=365 \rm \mu G$ for $R_{\rm out} = 0.1R_{\rm A}$ (see Figure \ref{fig:diff r_out}). 
\citet{2016ApJ...833...30C} pointed out that the magnetic field can be substantially amplified from the Bondi radius to the  circularization radius due to the magnetic freezing effect,
\begin{equation}
B(R_{\rm out}) \simeq \left(\frac{R_{\rm A}}{R_{\rm out}}\right)^2 B_{\rm A},
\end{equation}
 {where $B_{\rm A}$ is the magnetic field strength at the Bondi radius, which implies that a weaker field of the ISM at the Bondi radius is needed for the same jet power if the outer radius of the accretion flow is much smaller than $0.1~R_{\rm A}$.  }

\section{Discussion}  \label{sec_discuss}

It is still quite uncertain whether the accretion flow is MAD in M87, though alternative SANE (Standard and Normal Evolution) model is not completely ruled out \citep[e.g.,][]{2019MNRAS.486.2873C,2022ApJ...924..124Y}. 
\citet{2022ApJ...924..124Y} used the three-dimensional general relativistic magnetohydrodynamics (GRMHD) numerical simulation code ATHENA++ to simulate two modes of BH accretion flow, namely SANE and MAD. They compared the simulated predictions with the Faraday rotation measurement (RM) data of M87 \citep{2019ApJ...871..257P}. The RM value predicted by the MAD model is highly consistent with the observed data, while  The results of the SANE model is more than two orders of magnitude higher than the observed value. \cite{2019MNRAS.486.2873C} employed the MAD model and use the KORAL code to conduct three-dimensional GRMHD simulations, successfully reproducing the multiband observational features of the M87 jet, especially the high jet power. Most of these works only provided somewhat indirect evidence of MAD in M87, which is understandable because the a MAD has similar spectrum to that of the SANE disc\citep[][]{2019ApJ...887..167X}.

The BZ mechanism is believed to be responsible for jet formation, and therefore the field strength at the BH horizon can be estimated with the observed jet power \citep[][]{2015MNRAS.451..927Z}. MAD state can be maintained with either strong magnetic field or low surface density of the disc, or both, i.e., the ratio $B^2/\Sigma$ should be greater than a critical value (see Equation \ref{eq:MAD}).  On the other hand, the accretion rate in the inner region of the accretion flow in M87 can be well constrained by multi-waveband continuum spectrum and the RM data \citep[][]{2016ApJ...830....6F} , with which the surface density of the disc is derived. 
As the field can only be maintained by the currents in the accretion disc surrounding the BH \citep{1999ApJ...512..100L}, the field strength at the inner edge of the accretion flow is calculated by assuming the field of the disc is formed through dragging the external weak field inwards by the disc.  Our results show that the field is substantially enhanced in the accretion flow (from several hundred $\mu \rm G$ at $R_{\rm out}$ to hundreds G at $R_{\rm h}$, see Figure \ref{fig:BZ}) .  

The typical magnetic field strength of galaxy cluster atmospheres is of the order of $\sim \mu\rm G$ \citep[see][see references therein]{2002ARA&A..40..319C}, and the field strength of the extended halo of M87 is $\sim 10\mu\rm G$ \citep[][]{2012A&A...547A..56D}. It was pointed out that the field can be substantially amplified from the Bondi radius to the circularization radius by a a factor of $\sim (R_{\rm A}/R_{\rm out})^2$ due to field freezing effort \citep{2016ApJ...833...30C}. In the central region of our Galaxy, the field strength of the gas can be as high as $\sim \rm mG$ \citep[][]{2007A&A...464..609H}. The external field strength required in our calculations is roughly consistent with the these measurements. 

If the external magnetic field is stronger than a critical value, the inner edge of the ADAF is magnetically arrested, otherwise it is a SANE disc.  In this work, we adopt the jet power estimated with a radiative jet model \citep[][]{2015MNRAS.451..927Z}, which is instant jet power,  and is more suitable for our present investigation than those estimates of long-time averaged jet power (see Section \ref{sec_pjet_m87} for detailed discussion).  It is found that, in order to produce the observed jet power $P_{\rm jet}=3.2\times 10^{43}~\rm erg~s^{-1}$, the accretion flow in M87 must be magnetically arrested (see Figure \ref{fig:P_Bz}).  

We note that our conclusion depends sensitively on the measurement of jet power. If the jet power is significantly lower than $\sim 1.75 \times 10^{43} \rm erg~s^{-1}$, a SANE disc is allowed in M87. Thus, accurate measurement of instant jet power is the key point to resolve this problem.  The structure (i.e., the surface density) of the disc in the inner region is derived with the spectral fitting, in which a SANE model is adopted \citep{2016ApJ...830....6F}. We believe it will not affect our main conclusion on MAD, because the detailed calculations show the spectral difference between MAD and SANE models is observationally indistinguishable \citep{2019ApJ...887..167X}.

\section*{Acknowledgements}
We are grateful to the referee for the very helpful comments/suggestions. 
We thank Andrzej A. Zdziarski for helpful discussion. This work is supported by the NSFC (12533005, 12233007, 12347103, 12303020, 12563002 and 12361131579), the science research grants from the China Manned Space Project with CMS-CSST-2025-A07, and the fundamental research fund for Chinese central universities (Zhejiang University), the Yunnan fundamental research projects (NO.202401CF070169), the Xingdian Talent Support Plan-Youth Project.

\section*{Data Availability}

The data underlying this article will be shared on reasonable request to the corresponding author.



\bibliographystyle{mnras}
\bibliography{M87-MAD} 

@ARTICLE{2012ApJ...758...94P,
       author = {{Pellegrini}, Silvia and {Wang}, Junfeng and {Fabbiano}, Giuseppina and {Kim}, Dong-Woo and {Brassington}, Nicola J. and {Gallagher}, John S. and {Trinchieri}, Ginevra and {Zezas}, Andreas},
        title = "{AGN Activity and the Misaligned Hot ISM in the Compact Radio Elliptical NGC 4278}",
      journal = {\apj},
         year = 2012,
        month = oct,
       volume = {758},
       number = {2},
          eid = {94},
        pages = {94},
          doi = {10.1088/0004-637X/758/2/94},
archivePrefix = {arXiv},
       eprint = {1206.2533},
 primaryClass = {astro-ph.HE},
       adsurl = {https://ui.adsabs.harvard.edu/abs/2012ApJ...758...94P}
}

@ARTICLE{2004ApJ...613..322Q,
       author = {{Quataert}, Eliot},
        title = "{A Dynamical Model for Hot Gas in the Galactic Center}",
      journal = {\apj},
         year = 2004,
        month = sep,
       volume = {613},
       number = {1},
        pages = {322-325},
          doi = {10.1086/422973},
archivePrefix = {arXiv},
       eprint = {astro-ph/0310446},
 primaryClass = {astro-ph},
       adsurl = {https://ui.adsabs.harvard.edu/abs/2004ApJ...613..322Q}
}

@ARTICLE{1991ApJ...382L..19K,
       author = {{Krabbe}, A. and {Genzel}, R. and {Drapatz}, S. and {Rotaciuc}, V.},
        title = "{A Cluster of He i Emission-Line Stars in the Galactic Center}",
      journal = {\apjl},
         year = 1991,
        month = nov,
       volume = {382},
        pages = {L19},
          doi = {10.1086/186204},
       adsurl = {https://ui.adsabs.harvard.edu/abs/1991ApJ...382L..19K}
}

@ARTICLE{1992ApJ...387L..25M,
       author = {{Melia}, Fulvio},
        title = "{An Accreting Black Hole Model for Sagittarius A *}",
      journal = {\apjl},
         year = 1992,
        month = mar,
       volume = {387},
        pages = {L25},
          doi = {10.1086/186297},
       adsurl = {https://ui.adsabs.harvard.edu/abs/1992ApJ...387L..25M}
}

@ARTICLE{1997A&A...325..700N,
       author = {{Najarro}, F. and {Krabbe}, A. and {Genzel}, R. and {Lutz}, D. and {Kudritzki}, R.~P. and {Hillier}, D.~J.},
        title = "{Quantitative spectroscopy of the HeI cluster in the Galactic center.}",
      journal = {\aap},
         year = 1997,
        month = sep,
       volume = {325},
        pages = {700-708},
       adsurl = {https://ui.adsabs.harvard.edu/abs/1997A&A...325..700N}
}

@ARTICLE{2024ApJ...972...34L,
       author = {{Li}, Shuang-Liang and {Zuo}, Wenwen and {Cao}, Xinwu},
        title = "{Do All the Quasars and High-excitation Radio Galaxies (HERGs) in the 3CRR Catalog Contain a Magnetically Arrested Disk (MAD)?}",
      journal = {\apj},
         year = 2024,
        month = sep,
       volume = {972},
       number = {1},
          eid = {34},
        pages = {34},
          doi = {10.3847/1538-4357/ad6a5b},
archivePrefix = {arXiv},
       eprint = {2408.00321},
 primaryClass = {astro-ph.GA},
       adsurl = {https://ui.adsabs.harvard.edu/abs/2024ApJ...972...34L}
}

@BOOK{1979cmft.book.....P,
       author = {{Parker}, E.~N.},
        title = "{Cosmical magnetic fields. Their origin and their activity}",
         year = 1979,
       adsurl = {https://ui.adsabs.harvard.edu/abs/1979cmft.book.....P}
}

@INPROCEEDINGS{2015ASSL..414...45T,
       author = {{Tchekhovskoy}, Alexander},
        title = "{Launching of Active Galactic Nuclei Jets}",
    booktitle = {The Formation and Disruption of Black Hole Jets},
         year = 2015,
       editor = {{Contopoulos}, Ioannis and {Gabuzda}, Denise and {Kylafis}, Nikolaos},
       series = {Astrophysics and Space Science Library},
       volume = {414},
        month = jan,
        pages = {45},
          doi = {10.1007/978-3-319-10356-3_3},
       adsurl = {https://ui.adsabs.harvard.edu/abs/2015ASSL..414...45T}
}

@ARTICLE{2026A&A...708A.192G,
       author = {{Glaser}, Felix and {Fromm}, Christian M. and {Mizuno}, Yosuke and {Kadler}, Matthias and {Mannheim}, Karl},
        title = "{The magnetic filling in magnetically arrested accretion disk simulations and its impact on the jet in M87}",
      journal = {\aap},
         year = 2026,
        month = apr,
       volume = {708},
          eid = {A192},
        pages = {A192},
          doi = {10.1051/0004-6361/202555594},
       adsurl = {https://ui.adsabs.harvard.edu/abs/2026A&A...708A.192G}
}

@ARTICLE{2026A&A...705A.156S,
       author = {{Saiz-P{\'e}rez}, Ainara and {Fromm}, Christian M. and {Mizuno}, Yosuke and {Kadler}, Matthias and {Mannheim}, Karl and {Younsi}, Ziri},
        title = "{Probing the disk-jet coupling in M 87}",
      journal = {\aap},
         year = 2026,
        month = jan,
       volume = {705},
          eid = {A156},
        pages = {A156},
          doi = {10.1051/0004-6361/202556060},
archivePrefix = {arXiv},
       eprint = {2511.15482},
 primaryClass = {astro-ph.HE},
       adsurl = {https://ui.adsabs.harvard.edu/abs/2026A&A...705A.156S}
}

@ARTICLE{2022ApJ...924..124Y,
       author = {{Yuan}, Feng and {Wang}, Haiyang and {Yang}, Hai},
        title = "{The Accretion Flow in M87 is Really MAD}",
      journal = {\apj},
         year = 2022,
        month = jan,
       volume = {924},
       number = {2},
          eid = {124},
        pages = {124},
          doi = {10.3847/1538-4357/ac4714},
archivePrefix = {arXiv},
       eprint = {2201.00512},
 primaryClass = {astro-ph.HE},
       adsurl = {https://ui.adsabs.harvard.edu/abs/2022ApJ...924..124Y}
}

@ARTICLE{2019MNRAS.485.1916C,
       author = {{Cao}, Xinwu and {Lai}, Dong},
        title = "{Jet production in black hole X-ray binaries and active galactic nuclei: mass feeding and advection of magnetic fields}",
      journal = {\mnras},
         year = 2019,
        month = may,
       volume = {485},
       number = {2},
        pages = {1916-1923},
          doi = {10.1093/mnras/stz580},
archivePrefix = {arXiv},
       eprint = {1712.09265},
 primaryClass = {astro-ph.HE},
       adsurl = {https://ui.adsabs.harvard.edu/abs/2019MNRAS.485.1916C}
}

@ARTICLE{2023ApJ...944..182D,
       author = {{Dhang}, Prasun and {Bai}, Xue-Ning and {White}, Christopher J.},
        title = "{Magnetic Flux Transport in Radiatively Inefficient Accretion Flows and the Pathway toward a Magnetically Arrested Disk}",
      journal = {\apj},
         year = 2023,
        month = feb,
       volume = {944},
       number = {2},
          eid = {182},
        pages = {182},
          doi = {10.3847/1538-4357/acb534},
archivePrefix = {arXiv},
       eprint = {2208.02269},
 primaryClass = {astro-ph.HE},
       adsurl = {https://ui.adsabs.harvard.edu/abs/2023ApJ...944..182D}
}

@ARTICLE{1974Ap&SS..28...45B,
       author = {{Bisnovatyi-Kogan}, G.~S. and {Ruzmaikin}, A.~A.},
        title = "{The Accretion of Matter by a Collapsing Star in the Presence of a Magnetic Field}",
      journal = {\apss},
         year = 1974,
        month = may,
       volume = {28},
       number = {1},
        pages = {45-59},
          doi = {10.1007/BF00642237},
       adsurl = {https://ui.adsabs.harvard.edu/abs/1974Ap&SS..28...45B}
}

@ARTICLE{2009MNRAS.396..984G,
       author = {{Gu}, Minfeng and {Cao}, Xinwu and {Jiang}, D.~R.},
        title = "{The bulk kinetic power of radio jets in active galactic nuclei}",
      journal = {\mnras},
         year = 2009,
        month = jun,
       volume = {396},
       number = {2},
        pages = {984-996},
          doi = {10.1111/j.1365-2966.2009.14758.x},
archivePrefix = {arXiv},
       eprint = {0903.1896},
 primaryClass = {astro-ph.GA},
       adsurl = {https://ui.adsabs.harvard.edu/abs/2009MNRAS.396..984G}
}

@ARTICLE{1981ApJ...243..700K,
       author = {{Konigl}, A.},
        title = "{Relativistic jets as X-ray and gamma-ray sources.}",
      journal = {\apj},
         year = 1981,
        month = feb,
       volume = {243},
        pages = {700-709},
          doi = {10.1086/158638},
       adsurl = {https://ui.adsabs.harvard.edu/abs/1981ApJ...243..700K}
}

@ARTICLE{1998ApJ...494..139J,
       author = {{Jiang}, D.~R. and {Cao}, Xinwu and {Hong}, Xiaoyu},
        title = "{The Inhomogeneous Jet Parameters in Active Galactic Nuclei}",
      journal = {\apj},
         year = 1998,
        month = feb,
       volume = {494},
       number = {1},
        pages = {139-149},
          doi = {10.1086/305182},
archivePrefix = {arXiv},
       eprint = {astro-ph/9711264},
 primaryClass = {astro-ph},
       adsurl = {https://ui.adsabs.harvard.edu/abs/1998ApJ...494..139J}
}

@ARTICLE{2024ApJ...976..214N,
       author = {{Nied{\'z}wiecki}, Andrzej and {Szanecki}, Micha{\l} and {Janiuk}, Agnieszka},
        title = "{Broadband Spectral Modeling of the M87 Nucleus}",
      journal = {\apj},
         year = 2024,
        month = dec,
       volume = {976},
       number = {2},
          eid = {214},
        pages = {214},
          doi = {10.3847/1538-4357/ad88e9},
archivePrefix = {arXiv},
       eprint = {2406.17200},
 primaryClass = {astro-ph.HE},
       adsurl = {https://ui.adsabs.harvard.edu/abs/2024ApJ...976..214N}
}

@ARTICLE{2002ARA&A..40..319C,
       author = {{Carilli}, C.~L. and {Taylor}, G.~B.},
        title = "{Cluster Magnetic Fields}",
      journal = {\araa},
         year = 2002,
        month = jan,
       volume = {40},
        pages = {319-348},
          doi = {10.1146/annurev.astro.40.060401.093852},
archivePrefix = {arXiv},
       eprint = {astro-ph/0110655},
 primaryClass = {astro-ph},
       adsurl = {https://ui.adsabs.harvard.edu/abs/2002ARA&A..40..319C}
}

@ARTICLE{2007A&A...464..609H,
       author = {{Han}, J.~L. and {Zhang}, J.~S.},
        title = "{The Galactic distribution of magnetic fields in molecular clouds and HII regions}",
      journal = {\aap},
         year = 2007,
        month = mar,
       volume = {464},
       number = {2},
        pages = {609-614},
          doi = {10.1051/0004-6361:20065801},
archivePrefix = {arXiv},
       eprint = {astro-ph/0611213},
 primaryClass = {astro-ph},
       adsurl = {https://ui.adsabs.harvard.edu/abs/2007A&A...464..609H}
}

@ARTICLE{2012A&A...547A..56D,
       author = {{de Gasperin}, F. and {Orr{\'u}}, E. and {Murgia}, M. and {Merloni}, A. and {Falcke}, H. and {Beck}, R. and {Beswick}, R. and {B{\^\i}rzan}, L. and {Bonafede}, A. and {Br{\"u}ggen}, M. and {Brunetti}, G. and {Chy{\.z}y}, K. and {Conway}, J. and {Croston}, J.~H. and {En{\ss}lin}, T. and {Ferrari}, C. and {Heald}, G. and {Heidenreich}, S. and {Jackson}, N. and {Macario}, G. and {McKean}, J. and {Miley}, G. and {Morganti}, R. and {Offringa}, A. and {Pizzo}, R. and {Rafferty}, D. and {R{\"o}ttgering}, H. and {Shulevski}, A. and {Steinmetz}, M. and {Tasse}, C. and {van der Tol}, S. and {van Driel}, W. and {van Weeren}, R.~J. and {van Zwieten}, J.~E. and {Alexov}, A. and {Anderson}, J. and {Asgekar}, A. and {Avruch}, M. and {Bell}, M. and {Bell}, M.~R. and {Bentum}, M. and {Bernardi}, G. and {Best}, P. and {Breitling}, F. and {Broderick}, J.~W. and {Butcher}, A. and {Ciardi}, B. and {Dettmar}, R.~J. and {Eisloeffel}, J. and {Frieswijk}, W. and {Gankema}, H. and {Garrett}, M. and {Gerbers}, M. and {Griessmeier}, J.~M. and {Gunst}, A.~W. and {Hassall}, T.~E. and {Hessels}, J. and {Hoeft}, M. and {Horneffer}, A. and {Karastergiou}, A. and {K{\"o}hler}, J. and {Koopman}, Y. and {Kuniyoshi}, M. and {Kuper}, G. and {Maat}, P. and {Mann}, G. and {Mevius}, M. and {Mulcahy}, D.~D. and {Munk}, H. and {Nijboer}, R. and {Noordam}, J. and {Paas}, H. and {Pandey}, M. and {Pandey}, V.~N. and {Polatidis}, A. and {Reich}, W. and {Schoenmakers}, A.~P. and {Sluman}, J. and {Smirnov}, O. and {Sobey}, C. and {Stappers}, B. and {Swinbank}, J. and {Tagger}, M. and {Tang}, Y. and {van Bemmel}, I. and {van Cappellen}, W. and {van Duin}, A.~P. and {van Haarlem}, M. and {van Leeuwen}, J. and {Vermeulen}, R. and {Vocks}, C. and {White}, S. and {Wise}, M. and {Wucknitz}, O. and {Zarka}, P.},
        title = "{M 87 at metre wavelengths: the LOFAR picture}",
      journal = {\aap},
         year = 2012,
        month = nov,
       volume = {547},
          eid = {A56},
        pages = {A56},
          doi = {10.1051/0004-6361/201220209},
archivePrefix = {arXiv},
       eprint = {1210.1346},
 primaryClass = {astro-ph.GA},
       adsurl = {https://ui.adsabs.harvard.edu/abs/2012A&A...547A..56D}
}

@ARTICLE{2019ApJ...887..167X,
       author = {{Xie}, Fu-Guo and {Zdziarski}, Andrzej A.},
        title = "{Radiative Properties of Magnetically Arrested Disks}",
      journal = {\apj},
         year = 2019,
        month = dec,
       volume = {887},
       number = {2},
          eid = {167},
        pages = {167},
          doi = {10.3847/1538-4357/ab5848},
archivePrefix = {arXiv},
       eprint = {1911.06439},
 primaryClass = {astro-ph.HE},
       adsurl = {https://ui.adsabs.harvard.edu/abs/2019ApJ...887..167X}
}

@ARTICLE{2016ApJ...833...30C,
       author = {{Cao}, Xinwu},
        title = "{On the Radio Dichotomy of Active Galactic Nuclei}",
      journal = {\apj},
         year = 2016,
        month = dec,
       volume = {833},
       number = {1},
          eid = {30},
        pages = {30},
          doi = {10.3847/1538-4357/833/1/30},
archivePrefix = {arXiv},
       eprint = {1610.04061},
 primaryClass = {astro-ph.HE},
       adsurl = {https://ui.adsabs.harvard.edu/abs/2016ApJ...833...30C}
}

@ARTICLE{1972ApJ...178..347B,
       author = {{Bardeen}, James M. and {Press}, William H. and {Teukolsky}, Saul A.},
        title = "{Rotating Black Holes: Locally Nonrotating Frames, Energy Extraction, and Scalar Synchrotron Radiation}",
      journal = {\apj},
         year = 1972,
        month = dec,
       volume = {178},
        pages = {347-370},
          doi = {10.1086/151796},
       adsurl = {https://ui.adsabs.harvard.edu/abs/1972ApJ...178..347B}
}

@ARTICLE{1999MNRAS.303L...1B,
       author = {{Blandford}, Roger D. and {Begelman}, Mitchell C.},
        title = "{On the fate of gas accreting at a low rate on to a black hole}",
      journal = {\mnras},
         year = 1999,
        month = feb,
       volume = {303},
       number = {1},
        pages = {L1-L5},
          doi = {10.1046/j.1365-8711.1999.02358.x},
archivePrefix = {arXiv},
       eprint = {astro-ph/9809083},
 primaryClass = {astro-ph},
       adsurl = {https://ui.adsabs.harvard.edu/abs/1999MNRAS.303L...1B}
}

@ARTICLE{2011ApJ...737...94C,
       author = {{Cao}, Xinwu},
        title = "{The Large-scale Magnetic Fields of Advection-dominated Accretion Flows}",
      journal = {\apj},
         year = 2011,
        month = aug,
       volume = {737},
       number = {2},
          eid = {94},
        pages = {94},
          doi = {10.1088/0004-637X/737/2/94},
archivePrefix = {arXiv},
       eprint = {1105.6142},
 primaryClass = {astro-ph.HE},
       adsurl = {https://ui.adsabs.harvard.edu/abs/2011ApJ...737...94C}
}

@ARTICLE{1994MNRAS.267..235L,
       author = {{Lubow}, S.~H. and {Papaloizou}, J.~C.~B. and {Pringle}, J.~E.},
        title = "{Magnetic field dragging in accretion discs}",
      journal = {\mnras},
         year = 1994,
        month = mar,
       volume = {267},
       number = {2},
        pages = {235-240},
          doi = {10.1093/mnras/267.2.235},
       adsurl = {https://ui.adsabs.harvard.edu/abs/1994MNRAS.267..235L}
}

@ARTICLE{2003PASJ...55L..69N,
       author = {{Narayan}, Ramesh and {Igumenshchev}, Igor V. and {Abramowicz}, Marek A.},
        title = "{Magnetically Arrested Disk: an Energetically Efficient Accretion Flow}",
      journal = {\pasj},
         year = 2003,
        month = dec,
       volume = {55},
        pages = {L69-L72},
          doi = {10.1093/pasj/55.6.L69},
archivePrefix = {arXiv},
       eprint = {astro-ph/0305029},
 primaryClass = {astro-ph},
       adsurl = {https://ui.adsabs.harvard.edu/abs/2003PASJ...55L..69N}
}

@ARTICLE{2016ApJ...817...71C,
       author = {{Cao}, Xinwu},
        title = "{An Accretion Disk-outflow Model for Hysteretic State Transition in X-Ray Binaries}",
      journal = {\apj},
         year = 2016,
        month = jan,
       volume = {817},
       number = {1},
          eid = {71},
        pages = {71},
          doi = {10.3847/0004-637X/817/1/71},
archivePrefix = {arXiv},
       eprint = {1512.00124},
 primaryClass = {astro-ph.HE},
       adsurl = {https://ui.adsabs.harvard.edu/abs/2016ApJ...817...71C}
}

@ARTICLE{2017MNRAS.470..612F,
       author = {{Feng}, Jianchao and {Wu}, Qingwen},
        title = "{Constraint on the black hole spin of M87 from the accretion-jet model}",
      journal = {\mnras},
         year = 2017,
        month = may,
       volume = {470},
       number = {1},
        pages = {612-616},
          doi = {10.1093/mnras/stx1283},
archivePrefix = {arXiv},
       eprint = {1705.07804},
 primaryClass = {astro-ph.HE},
       adsurl = {https://ui.adsabs.harvard.edu/abs/2017MNRAS.470..612F}
}

@ARTICLE{2016ApJ...830....6F,
       author = {{Feng}, Jianchao and {Wu}, Qingwen and {Lu}, Ru-Sen},
        title = "{An Accretion-jet Model for M87: Interpreting the Spectral Energy Distribution and Faraday Rotation Measure}",
      journal = {\apj},
         year = 2016,
        month = oct,
       volume = {830},
       number = {1},
          eid = {6},
        pages = {6},
          doi = {10.3847/0004-637X/830/1/6},
archivePrefix = {arXiv},
       eprint = {1607.08054},
 primaryClass = {astro-ph.HE},
       adsurl = {https://ui.adsabs.harvard.edu/abs/2016ApJ...830....6F}
}

@ARTICLE{2004ApJ...611..977M,
       author = {{McKinney}, Jonathan C. and {Gammie}, Charles F.},
        title = "{A Measurement of the Electromagnetic Luminosity of a Kerr Black Hole}",
      journal = {\apj},
         year = 2004,
        month = aug,
       volume = {611},
       number = {2},
        pages = {977-995},
          doi = {10.1086/422244},
archivePrefix = {arXiv},
       eprint = {astro-ph/0404512},
 primaryClass = {astro-ph},
       adsurl = {https://ui.adsabs.harvard.edu/abs/2004ApJ...611..977M}
}

@ARTICLE{2006MNRAS.372...21A,
       author = {{Allen}, S.~W. and {Dunn}, R.~J.~H. and {Fabian}, A.~C. and {Taylor}, G.~B. and {Reynolds}, C.~S.},
        title = "{The relation between accretion rate and jet power in X-ray luminous elliptical galaxies}",
      journal = {\mnras},
         year = 2006,
        month = oct,
       volume = {372},
       number = {1},
        pages = {21-30},
          doi = {10.1111/j.1365-2966.2006.10778.x},
archivePrefix = {arXiv},
       eprint = {astro-ph/0602549},
 primaryClass = {astro-ph},
       adsurl = {https://ui.adsabs.harvard.edu/abs/2006MNRAS.372...21A}
}

@ARTICLE{2006ApJ...652..216R,
       author = {{Rafferty}, D.~A. and {McNamara}, B.~R. and {Nulsen}, P.~E.~J. and {Wise}, M.~W.},
        title = "{The Feedback-regulated Growth of Black Holes and Bulges through Gas Accretion and Starbursts in Cluster Central Dominant Galaxies}",
      journal = {\apj},
         year = 2006,
        month = nov,
       volume = {652},
       number = {1},
        pages = {216-231},
          doi = {10.1086/507672},
archivePrefix = {arXiv},
       eprint = {astro-ph/0605323},
 primaryClass = {astro-ph},
       adsurl = {https://ui.adsabs.harvard.edu/abs/2006ApJ...652..216R}
}

@ARTICLE{2004ApJ...607..800B,
       author = {{B{\^\i}rzan}, L. and {Rafferty}, D.~A. and {McNamara}, B.~R. and {Wise}, M.~W. and {Nulsen}, P.~E.~J.},
        title = "{A Systematic Study of Radio-induced X-Ray Cavities in Clusters, Groups, and Galaxies}",
      journal = {\apj},
         year = 2004,
        month = jun,
       volume = {607},
       number = {2},
        pages = {800-809},
          doi = {10.1086/383519},
archivePrefix = {arXiv},
       eprint = {astro-ph/0402348},
 primaryClass = {astro-ph},
       adsurl = {https://ui.adsabs.harvard.edu/abs/2004ApJ...607..800B}
}

@ARTICLE{2015MNRAS.451..927Z,
       author = {{Zdziarski}, Andrzej A. and {Sikora}, Marek and {Pjanka}, Patryk and {Tchekhovskoy}, Alexander},
        title = "{Core shifts, magnetic fields and magnetization of extragalactic jets}",
      journal = {\mnras},
         year = 2015,
        month = jul,
       volume = {451},
       number = {1},
        pages = {927-935},
          doi = {10.1093/mnras/stv986},
archivePrefix = {arXiv},
       eprint = {1410.7310},
 primaryClass = {astro-ph.HE},
       adsurl = {https://ui.adsabs.harvard.edu/abs/2015MNRAS.451..927Z}
}

@ARTICLE{2023Sci...381..961Y,
       author = {{You}, Bei and {Cao}, Xinwu and {Yan}, Zhen and {Hameury}, Jean-Marie and {Czerny}, Bozena and {Wu}, Yue and {Xia}, Tianyu and {Sikora}, Marek and {Zhang}, Shuang-Nan and {Du}, Pu and {Zycki}, Piotr T.},
        title = "{Observations of a black hole x-ray binary indicate formation of a magnetically arrested disk}",
      journal = {Science},
         year = 2023,
        month = sep,
       volume = {381},
       number = {6661},
        pages = {961-964},
          doi = {10.1126/science.abo4504},
archivePrefix = {arXiv},
       eprint = {2309.00200},
 primaryClass = {astro-ph.HE},
       adsurl = {https://ui.adsabs.harvard.edu/abs/2023Sci...381..961Y}
}

@ARTICLE{1999ApJ...512..100L,
       author = {{Livio}, M. and {Ogilvie}, G.~I. and {Pringle}, J.~E.},
        title = "{Extracting Energy from Black Holes: The Relative Importance of the Blandford-Znajek Mechanism}",
      journal = {\apj},
         year = 1999,
        month = feb,
       volume = {512},
       number = {1},
        pages = {100-104},
          doi = {10.1086/306777},
archivePrefix = {arXiv},
       eprint = {astro-ph/9809093},
 primaryClass = {astro-ph},
       adsurl = {https://ui.adsabs.harvard.edu/abs/1999ApJ...512..100L}
}

@ARTICLE{2009MNRAS.400.1734L,
       author = {{Li}, Shuang-Liang and {Cao}, Xinwu},
        title = "{Global dynamics of advection-dominated accretion flows with magnetically driven outflow}",
      journal = {\mnras},
         year = 2009,
        month = dec,
       volume = {400},
       number = {4},
        pages = {1734-1741},
          doi = {10.1111/j.1365-2966.2009.15595.x},
archivePrefix = {arXiv},
       eprint = {0908.3370},
 primaryClass = {astro-ph.HE},
       adsurl = {https://ui.adsabs.harvard.edu/abs/2009MNRAS.400.1734L}
}

@ARTICLE{2009ApJ...694..556B,
       author = {{Blakeslee}, John P. and {Jord{\'a}n}, Andr{\'e}s and {Mei}, Simona and {C{\^o}t{\'e}}, Patrick and {Ferrarese}, Laura and {Infante}, Leopoldo and {Peng}, Eric W. and {Tonry}, John L. and {West}, Michael J.},
        title = "{The ACS Fornax Cluster Survey. V. Measurement and Recalibration of Surface Brightness Fluctuations and a Precise Value of the Fornax-Virgo Relative Distance}",
      journal = {\apj},
         year = 2009,
        month = mar,
       volume = {694},
       number = {1},
        pages = {556-572},
          doi = {10.1088/0004-637X/694/1/556},
archivePrefix = {arXiv},
       eprint = {0901.1138},
 primaryClass = {astro-ph.CO},
       adsurl = {https://ui.adsabs.harvard.edu/abs/2009ApJ...694..556B}
}

@ARTICLE{2021ApJ...911L..11E,
       author = {{EHT MWL Science Working Group} and {Algaba}, J.~C. and {Anczarski}, J. and {Asada}, K. and {Balokovi{\'c}}, M. and {Chandra}, S. and {Cui}, Y. -Z. and {Falcone}, A.~D. and {Giroletti}, M. and {Goddi}, C. and {Hada}, K. and {Haggard}, D. and {Jorstad}, S. and {Kaur}, A. and {Kawashima}, T. and {Keating}, G. and {Kim}, J. -Y. and {Kino}, M. and {Komossa}, S. and {Kravchenko}, E.~V. and {Krichbaum}, T.~P. and {Lee}, S. -S. and {Lu}, R. -S. and {Lucchini}, M. and {Markoff}, S. and {Neilsen}, J. and {Nowak}, M.~A. and {Park}, J. and {Principe}, G. and {Ramakrishnan}, V. and {Reynolds}, M.~T. and {Sasada}, M. and {Savchenko}, S.~S. and {Williamson}, K.~E. and {Event Horizon Telescope Collaboration} and {Akiyama}, Kazunori and {Alberdi}, Antxon and {Alef}, Walter and {Anantua}, Richard and {Azulay}, Rebecca and {Baczko}, Anne-Kathrin and {Ball}, David and {Barrett}, John and {Bintley}, Dan and {Benson}, Bradford A. and {Blackburn}, Lindy and {Blundell}, Raymond and {Boland}, Wilfred and {Bouman}, Katherine L. and {Bower}, Geoffrey C. and {Boyce}, Hope and {Bremer}, Michael and {Brinkerink}, Christiaan D. and {Brissenden}, Roger and {Britzen}, Silke and {Broderick}, Avery E. and {Broguiere}, Dominique and {Bronzwaer}, Thomas and {Byun}, Do-Young and {Carlstrom}, John E. and {Chael}, Andrew and {Chan}, Chi-Kwan and {Chatterjee}, Shami and {Chatterjee}, Koushik and {Chen}, Ming-Tang and {Chen}, Yongjun and {Chesler}, Paul M. and {Cho}, Ilje and {Christian}, Pierre and {Conway}, John E. and {Cordes}, James M. and {Crawford}, Thomas M. and {Crew}, Geoffrey B. and {Cruz-Osorio}, Alejandro and {Davelaar}, Jordy and {de Laurentis}, Mariafelicia and {Deane}, Roger and {Dempsey}, Jessica and {Desvignes}, Gregory and {Dexter}, Jason and {Doeleman}, Sheperd S. and {Eatough}, Ralph P. and {Falcke}, Heino and {Farah}, Joseph and {Fish}, Vincent L. and {Fomalont}, Ed and {Ford}, H. Alyson and {Fraga-Encinas}, Raquel and {Friberg}, Per and {Fromm}, Christian M. and {Fuentes}, Antonio and {Galison}, Peter and {Gammie}, Charles F. and {Garc{\'\i}a}, Roberto and {Gentaz}, Olivier and {Georgiev}, Boris and {Gold}, Roman and {G{\'o}mez}, Jos{\'e} L. and {G{\'o}mez-Ruiz}, Arturo I. and {Gu}, Minfeng and {Gurwell}, Mark and {Hecht}, Michael H. and {Hesper}, Ronald and {Ho}, Luis C. and {Ho}, Paul and {Honma}, Mareki and {Huang}, Chih-Wei L. and {Huang}, Lei and {Hughes}, David H. and {Ikeda}, Shiro and {Inoue}, Makoto and {Issaoun}, Sara and {James}, David J. and {Jannuzi}, Buell T. and {Janssen}, Michael and {Jeter}, Britton and {Jiang}, Wu and {Jim{\'e}nez-Rosales}, Alejandra and {Johnson}, Michael D. and {Jung}, Taehyun and {Karami}, Mansour and {Karuppusamy}, Ramesh and {Kettenis}, Mark and {Kim}, Dong-Jin and {Kim}, Jongsoo and {Kim}, Junhan and {Koay}, Jun Yi and {Kofuji}, Yutaro and {Koch}, Patrick M. and {Koyama}, Shoko and {Kramer}, Michael and {Kramer}, Carsten and {Kuo}, Cheng-Yu and {Lauer}, Tod R. and {Levis}, Aviad and {Li}, Yan-Rong and {Li}, Zhiyuan and {Lindqvist}, Michael and {Lico}, Rocco and {Lindahl}, Greg and {Liu}, Jun and {Liu}, Kuo and {Liuzzo}, Elisabetta and {Lo}, Wen-Ping and {Lobanov}, Andrei P. and {Loinard}, Laurent and {Lonsdale}, Colin and {MacDonald}, Nicholas R. and {Mao}, Jirong and {Marchili}, Nicola and {Marrone}, Daniel P. and {Marscher}, Alan P. and {Mart{\'\i}-Vidal}, Iv{\'a}n and {Matsushita}, Satoki and {Matthews}, Lynn D. and {Medeiros}, Lia and {Menten}, Karl M. and {Mizuno}, Izumi and {Mizuno}, Yosuke and {Moran}, James M. and {Moriyama}, Kotaro and {Moscibrodzka}, Monika and {M{\"u}ller}, Cornelia and {Musoke}, Gibwa and {Mej{\'\i}as}, Alejandro Mus and {Nagai}, Hiroshi and {Nagar}, Neil M. and {Nakamura}, Masanori and {Narayan}, Ramesh and {Narayanan}, Gopal and {Natarajan}, Iniyan and {Nathanail}, Antonios and {Neri}, Roberto and {Ni}, Chunchong and {Noutsos}, Aristeidis and {Okino}, Hiroki and {Olivares}, H{\'e}ctor and {Ortiz-Le{\'o}n}, Gisela N. and {Oyama}, Tomoaki and {{\"O}zel}, Feryal and {Palumbo}, Daniel C.~M. and {Patel}, Nimesh and {Pen}, Ue-Li and {Pesce}, Dominic W. and {Pi{\'e}tu}, Vincent and {Plambeck}, Richard and {Popstefanija}, Aleksandar and {Porth}, Oliver and {P{\"o}tzl}, Felix M. and {Prather}, Ben and {Preciado-L{\'o}pez}, Jorge A. and {Psaltis}, Dimitrios and {Pu}, Hung-Yi and {Rao}, Ramprasad and {Rawlings}, Mark G. and {Raymond}, Alexander W. and {Rezzolla}, Luciano and {Ricarte}, Angelo and {Ripperda}, Bart and {Roelofs}, Freek},
        title = "{Broadband Multi-wavelength Properties of M87 during the 2017 Event Horizon Telescope Campaign}",
      journal = {\apjl},
         year = 2021,
        month = apr,
       volume = {911},
       number = {1},
          eid = {L11},
        pages = {L11},
          doi = {10.3847/2041-8213/abef71},
archivePrefix = {arXiv},
       eprint = {2104.06855},
 primaryClass = {astro-ph.HE},
       adsurl = {https://ui.adsabs.harvard.edu/abs/2021ApJ...911L..11E}
}

@ARTICLE{2019ApJ...875L...6E,
       author = {{Event Horizon Telescope Collaboration} and {Akiyama}, Kazunori and {Alberdi}, Antxon and {Alef}, Walter and {Asada}, Keiichi and {Azulay}, Rebecca and {Baczko}, Anne-Kathrin and {Ball}, David and {Balokovi{\'c}}, Mislav and {Barrett}, John and {Bintley}, Dan and {Blackburn}, Lindy and {Boland}, Wilfred and {Bouman}, Katherine L. and {Bower}, Geoffrey C. and {Bremer}, Michael and {Brinkerink}, Christiaan D. and {Brissenden}, Roger and {Britzen}, Silke and {Broderick}, Avery E. and {Broguiere}, Dominique and {Bronzwaer}, Thomas and {Byun}, Do-Young and {Carlstrom}, John E. and {Chael}, Andrew and {Chan}, Chi-kwan and {Chatterjee}, Shami and {Chatterjee}, Koushik and {Chen}, Ming-Tang and {Chen}, Yongjun and {Cho}, Ilje and {Christian}, Pierre and {Conway}, John E. and {Cordes}, James M. and {Crew}, Geoffrey B. and {Cui}, Yuzhu and {Davelaar}, Jordy and {De Laurentis}, Mariafelicia and {Deane}, Roger and {Dempsey}, Jessica and {Desvignes}, Gregory and {Dexter}, Jason and {Doeleman}, Sheperd S. and {Eatough}, Ralph P. and {Falcke}, Heino and {Fish}, Vincent L. and {Fomalont}, Ed and {Fraga-Encinas}, Raquel and {Friberg}, Per and {Fromm}, Christian M. and {G{\'o}mez}, Jos{\'e} L. and {Galison}, Peter and {Gammie}, Charles F. and {Garc{\'\i}a}, Roberto and {Gentaz}, Olivier and {Georgiev}, Boris and {Goddi}, Ciriaco and {Gold}, Roman and {Gu}, Minfeng and {Gurwell}, Mark and {Hada}, Kazuhiro and {Hecht}, Michael H. and {Hesper}, Ronald and {Ho}, Luis C. and {Ho}, Paul and {Honma}, Mareki and {Huang}, Chih-Wei L. and {Huang}, Lei and {Hughes}, David H. and {Ikeda}, Shiro and {Inoue}, Makoto and {Issaoun}, Sara and {James}, David J. and {Jannuzi}, Buell T. and {Janssen}, Michael and {Jeter}, Britton and {Jiang}, Wu and {Johnson}, Michael D. and {Jorstad}, Svetlana and {Jung}, Taehyun and {Karami}, Mansour and {Karuppusamy}, Ramesh and {Kawashima}, Tomohisa and {Keating}, Garrett K. and {Kettenis}, Mark and {Kim}, Jae-Young and {Kim}, Junhan and {Kim}, Jongsoo and {Kino}, Motoki and {Koay}, Jun Yi and {Koch}, Patrick M. and {Koyama}, Shoko and {Kramer}, Michael and {Kramer}, Carsten and {Krichbaum}, Thomas P. and {Kuo}, Cheng-Yu and {Lauer}, Tod R. and {Lee}, Sang-Sung and {Li}, Yan-Rong and {Li}, Zhiyuan and {Lindqvist}, Michael and {Liu}, Kuo and {Liuzzo}, Elisabetta and {Lo}, Wen-Ping and {Lobanov}, Andrei P. and {Loinard}, Laurent and {Lonsdale}, Colin and {Lu}, Ru-Sen and {MacDonald}, Nicholas R. and {Mao}, Jirong and {Markoff}, Sera and {Marrone}, Daniel P. and {Marscher}, Alan P. and {Mart{\'\i}-Vidal}, Iv{\'a}n and {Matsushita}, Satoki and {Matthews}, Lynn D. and {Medeiros}, Lia and {Menten}, Karl M. and {Mizuno}, Yosuke and {Mizuno}, Izumi and {Moran}, James M. and {Moriyama}, Kotaro and {Moscibrodzka}, Monika and {M{\"u}ller}, Cornelia and {Nagai}, Hiroshi and {Nagar}, Neil M. and {Nakamura}, Masanori and {Narayan}, Ramesh and {Narayanan}, Gopal and {Natarajan}, Iniyan and {Neri}, Roberto and {Ni}, Chunchong and {Noutsos}, Aristeidis and {Okino}, Hiroki and {Olivares}, H{\'e}ctor and {Oyama}, Tomoaki and {{\"O}zel}, Feryal and {Palumbo}, Daniel C.~M. and {Patel}, Nimesh and {Pen}, Ue-Li and {Pesce}, Dominic W. and {Pi{\'e}tu}, Vincent and {Plambeck}, Richard and {PopStefanija}, Aleksandar and {Porth}, Oliver and {Prather}, Ben and {Preciado-L{\'o}pez}, Jorge A. and {Psaltis}, Dimitrios and {Pu}, Hung-Yi and {Ramakrishnan}, Venkatessh and {Rao}, Ramprasad and {Rawlings}, Mark G. and {Raymond}, Alexander W. and {Rezzolla}, Luciano and {Ripperda}, Bart and {Roelofs}, Freek and {Rogers}, Alan and {Ros}, Eduardo and {Rose}, Mel and {Roshanineshat}, Arash and {Rottmann}, Helge and {Roy}, Alan L. and {Ruszczyk}, Chet and {Ryan}, Benjamin R. and {Rygl}, Kazi L.~J. and {S{\'a}nchez}, Salvador and {S{\'a}nchez-Arguelles}, David and {Sasada}, Mahito and {Savolainen}, Tuomas and {Schloerb}, F. Peter and {Schuster}, Karl-Friedrich and {Shao}, Lijing and {Shen}, Zhiqiang and {Small}, Des and {Sohn}, Bong Won and {SooHoo}, Jason and {Tazaki}, Fumie and {Tiede}, Paul and {Tilanus}, Remo P.~J. and {Titus}, Michael and {Toma}, Kenji and {Torne}, Pablo and {Trent}, Tyler and {Trippe}, Sascha and {Tsuda}, Shuichiro and {van Bemmel}, Ilse and {van Langevelde}, Huib Jan and {van Rossum}, Daniel R. and {Wagner}, Jan and {Wardle}, John and {Weintroub}, Jonathan and {Wex}, Norbert and {Wharton}, Robert and {Wielgus}, Maciek and {Wong}, George N. and {Wu}, Qingwen and {Young}, Andr{\'e} and {Young}, Ken and {Younsi}, Ziri and {Yuan}, Feng},
        title = "{First M87 Event Horizon Telescope Results. VI. The Shadow and Mass of the Central Black Hole}",
      journal = {\apjl},
         year = 2019,
        month = apr,
       volume = {875},
       number = {1},
          eid = {L6},
        pages = {L6},
          doi = {10.3847/2041-8213/ab1141},
archivePrefix = {arXiv},
       eprint = {1906.11243},
 primaryClass = {astro-ph.GA},
       adsurl = {https://ui.adsabs.harvard.edu/abs/2019ApJ...875L...6E}
}

@ARTICLE{2018ApJ...855..128W,
       author = {{Walker}, R. Craig and {Hardee}, Philip E. and {Davies}, Frederick B. and {Ly}, Chun and {Junor}, William},
        title = "{The Structure and Dynamics of the Subparsec Jet in M87 Based on 50 VLBA Observations over 17 Years at 43 GHz}",
      journal = {\apj},
         year = 2018,
        month = mar,
       volume = {855},
       number = {2},
          eid = {128},
        pages = {128},
          doi = {10.3847/1538-4357/aaafcc},
archivePrefix = {arXiv},
       eprint = {1802.06166},
 primaryClass = {astro-ph.HE},
       adsurl = {https://ui.adsabs.harvard.edu/abs/2018ApJ...855..128W}
}

@ARTICLE{2002ApJ...579..560Y,
       author = {{Young}, A.~J. and {Wilson}, A.~S. and {Mundell}, C.~G.},
        title = "{Chandra Imaging of the X-Ray Core of the Virgo Cluster}",
      journal = {\apj},
         year = 2002,
        month = nov,
       volume = {579},
       number = {2},
        pages = {560-570},
          doi = {10.1086/342918},
archivePrefix = {arXiv},
       eprint = {astro-ph/0202504},
 primaryClass = {astro-ph},
       adsurl = {https://ui.adsabs.harvard.edu/abs/2002ApJ...579..560Y}
}

@ARTICLE{2014ARA&A..52..529Y,
       author = {{Yuan}, Feng and {Narayan}, Ramesh},
        title = "{Hot Accretion Flows Around Black Holes}",
      journal = {\araa},
         year = 2014,
        month = aug,
       volume = {52},
        pages = {529-588},
          doi = {10.1146/annurev-astro-082812-141003},
archivePrefix = {arXiv},
       eprint = {1401.0586},
 primaryClass = {astro-ph.HE},
       adsurl = {https://ui.adsabs.harvard.edu/abs/2014ARA&A..52..529Y}
}

@ARTICLE{2019ApJ...871..257P,
       author = {{Park}, Jongho and {Hada}, Kazuhiro and {Kino}, Motoki and {Nakamura}, Masanori and {Ro}, Hyunwook and {Trippe}, Sascha},
        title = "{Faraday Rotation in the Jet of M87 inside the Bondi Radius: Indication of Winds from Hot Accretion Flows Confining the Relativistic Jet}",
      journal = {\apj},
         year = 2019,
        month = feb,
       volume = {871},
       number = {2},
          eid = {257},
        pages = {257},
          doi = {10.3847/1538-4357/aaf9a9},
archivePrefix = {arXiv},
       eprint = {1812.08386},
 primaryClass = {astro-ph.HE},
       adsurl = {https://ui.adsabs.harvard.edu/abs/2019ApJ...871..257P}
}

@ARTICLE{2019MNRAS.486.2873C,
       author = {{Chael}, Andrew and {Narayan}, Ramesh and {Johnson}, Michael D.},
        title = "{Two-temperature, Magnetically Arrested Disc simulations of the jet from the supermassive black hole in M87}",
      journal = {\mnras},
         year = 2019,
        month = jun,
       volume = {486},
       number = {2},
        pages = {2873-2895},
          doi = {10.1093/mnras/stz988},
archivePrefix = {arXiv},
       eprint = {1810.01983},
 primaryClass = {astro-ph.HE},
       adsurl = {https://ui.adsabs.harvard.edu/abs/2019MNRAS.486.2873C}
}

@ARTICLE{1952MNRAS.112..195B,
       author = {{Bondi}, H.},
        title = "{On spherically symmetrical accretion}",
      journal = {\mnras},
         year = 1952,
        month = jan,
       volume = {112},
        pages = {195},
          doi = {10.1093/mnras/112.2.195},
       adsurl = {https://ui.adsabs.harvard.edu/abs/1952MNRAS.112..195B}
}

@ARTICLE{2008MNRAS.383..458C,
       author = {{Cuadra}, Jorge and {Nayakshin}, Sergei and {Martins}, Fabrice},
        title = "{Variable accretion and emission from the stellar winds in the Galactic Centre}",
      journal = {\mnras},
         year = 2008,
        month = jan,
       volume = {383},
       number = {2},
        pages = {458-466},
          doi = {10.1111/j.1365-2966.2007.12573.x},
archivePrefix = {arXiv},
       eprint = {0705.0769},
 primaryClass = {astro-ph},
       adsurl = {https://ui.adsabs.harvard.edu/abs/2008MNRAS.383..458C}
}

@ARTICLE{2014ApJ...783L..33K,
       author = {{Kuo}, C.~Y. and {Asada}, K. and {Rao}, R. and {Nakamura}, M. and {Algaba}, J.~C. and {Liu}, H.~B. and {Inoue}, M. and {Koch}, P.~M. and {Ho}, P.~T.~P. and {Matsushita}, S. and {Pu}, H.-Y. and {Akiyama}, K. and {Nishioka}, H. and {Pradel}, N.},
        title = "{Measuring Mass Accretion Rate onto the Supermassive Black Hole in M87 Using Faraday Rotation Measure with the Submillimeter Array}",
      journal = {\apjl},
         year = 2014,
        month = mar,
       volume = {783},
       number = {2},
          eid = {L33},
        pages = {L33},
          doi = {10.1088/2041-8205/783/2/L33},
archivePrefix = {arXiv},
       eprint = {1402.5238},
 primaryClass = {astro-ph.GA},
       adsurl = {https://ui.adsabs.harvard.edu/abs/2014ApJ...783L..33K}
}

@ARTICLE{1995ApJ...452..710N,
       author = {{Narayan}, Ramesh and {Yi}, Insu},
        title = "{Advection-dominated Accretion: Underfed Black Holes and Neutron Stars}",
      journal = {\apj},
         year = 1995,
        month = oct,
       volume = {452},
        pages = {710},
          doi = {10.1086/176343},
archivePrefix = {arXiv},
       eprint = {astro-ph/9411059},
 primaryClass = {astro-ph},
       adsurl = {https://ui.adsabs.harvard.edu/abs/1995ApJ...452..710N}
}

@ARTICLE{1977MNRAS.179..433B,
       author = {{Blandford}, R.~D. and {Znajek}, R.~L.},
        title = "{Electromagnetic extraction of energy from Kerr black holes.}",
      journal = {\mnras},
         year = 1977,
        month = may,
       volume = {179},
        pages = {433-456},
          doi = {10.1093/mnras/179.3.433},
       adsurl = {https://ui.adsabs.harvard.edu/abs/1977MNRAS.179..433B}
}

@ARTICLE{2024SciA...10N3544Y,
       author = {{Yang}, Hai and {Yuan}, Feng and {Li}, Hui and {Mizuno}, Yosuke and {Guo}, Fan and {Lu}, Rusen and {Ho}, Luis C. and {Lin}, Xi and {Zdziarski}, Andrzej A. and {Wang}, Jieshuang},
        title = "{Modeling the inner part of the jet in M87: Confronting jet morphology with theory}",
      journal = {Science Advances},
         year = 2024,
        month = mar,
       volume = {10},
       number = {12},
          eid = {eadn3544},
        pages = {eadn3544},
          doi = {10.1126/sciadv.adn3544},
archivePrefix = {arXiv},
       eprint = {2403.15950},
 primaryClass = {astro-ph.HE},
       adsurl = {https://ui.adsabs.harvard.edu/abs/2024SciA...10N3544Y}
}

@ARTICLE{2014ApJ...786....5K,
       author = {{Kino}, M. and {Takahara}, F. and {Hada}, K. and {Doi}, A.},
        title = "{Relativistic Electrons and Magnetic Fields of the M87 Jet on the \raisebox{-0.5ex}\textasciitilde10 Schwarzschild Radii Scale}",
      journal = {\apj},
         year = 2014,
        month = may,
       volume = {786},
       number = {1},
          eid = {5},
        pages = {5},
          doi = {10.1088/0004-637X/786/1/5},
archivePrefix = {arXiv},
       eprint = {1403.0650},
 primaryClass = {astro-ph.HE},
       adsurl = {https://ui.adsabs.harvard.edu/abs/2014ApJ...786....5K}
}

@ARTICLE{1979ApJ...232...34B,
       author = {{Blandford}, R.~D. and {K{\"o}nigl}, A.},
        title = "{Relativistic jets as compact radio sources.}",
      journal = {\apj},
         year = 1979,
        month = aug,
       volume = {232},
        pages = {34-48},
          doi = {10.1086/157262},
       adsurl = {https://ui.adsabs.harvard.edu/abs/1979ApJ...232...34B}
}

@ARTICLE{2014Natur.510..126Z,
       author = {{Zamaninasab}, M. and {Clausen-Brown}, E. and {Savolainen}, T. and {Tchekhovskoy}, A.},
        title = "{Dynamically important magnetic fields near accreting supermassive black holes}",
      journal = {\nat},
         year = 2014,
        month = jun,
       volume = {510},
       number = {7503},
        pages = {126-128},
          doi = {10.1038/nature13399},
       adsurl = {https://ui.adsabs.harvard.edu/abs/2014Natur.510..126Z}
}

@ARTICLE{2009A&A...507...19F,
       author = {{Fromang}, S. and {Stone}, J.~M.},
        title = "{Turbulent resistivity driven by the magnetorotational instability}",
      journal = {\aap},
         year = 2009,
        month = nov,
       volume = {507},
       number = {1},
        pages = {19-28},
          doi = {10.1051/0004-6361/200912752},
archivePrefix = {arXiv},
       eprint = {0906.4422},
 primaryClass = {astro-ph.SR},
       adsurl = {https://ui.adsabs.harvard.edu/abs/2009A&A...507...19F}
}

@ARTICLE{2009A&A...504..309L,
       author = {{Lesur}, G. and {Longaretti}, P.-Y.},
        title = "{Turbulent resistivity evaluation in magnetorotational instability generated turbulence}",
      journal = {\aap},
         year = 2009,
        month = sep,
       volume = {504},
       number = {2},
        pages = {309-320},
          doi = {10.1051/0004-6361/200912272},
archivePrefix = {arXiv},
       eprint = {0907.1393},
 primaryClass = {astro-ph.HE},
       adsurl = {https://ui.adsabs.harvard.edu/abs/2009A&A...504..309L}
}

@ARTICLE{2009ApJ...697.1901G,
       author = {{Guan}, Xiaoyue and {Gammie}, Charles F.},
        title = "{The Turbulent Magnetic Prandtl Number of MHD Turbulence in Disks}",
      journal = {\apj},
         year = 2009,
        month = jun,
       volume = {697},
       number = {2},
        pages = {1901-1906},
          doi = {10.1088/0004-637X/697/2/1901},
archivePrefix = {arXiv},
       eprint = {0903.3757},
 primaryClass = {astro-ph.HE},
       adsurl = {https://ui.adsabs.harvard.edu/abs/2009ApJ...697.1901G}
}

@ARTICLE{2003A&A...411..321Y,
       author = {{Yousef}, T.~A. and {Brandenburg}, A. and {R{\"u}diger}, G.},
        title = "{Turbulent magnetic Prandtl number and magnetic diffusivity quenching from simulations}",
      journal = {\aap},
         year = 2003,
        month = dec,
       volume = {411},
        pages = {321-327},
          doi = {10.1051/0004-6361:20031371},
archivePrefix = {arXiv},
       eprint = {astro-ph/0302425},
 primaryClass = {astro-ph},
       adsurl = {https://ui.adsabs.harvard.edu/abs/2003A&A...411..321Y}
}




\appendix




\bsp	
\label{lastpage}
\end{document}